\documentclass[twocolumn,superscriptaddress,prl,amsmath,amstex,amssymb,citeautoscript,longbibliography]{revtex4-1}
\pdfoutput=1
\usepackage{natbib}
\usepackage[english]{babel}
\usepackage{letltxmacro}
\usepackage{latexsym}
\usepackage[utf8]{inputenc}
\LetLtxMacro{\ORIGselectlanguage}{\selectlanguage}
\makeatletter
\DeclareRobustCommand{\selectlanguage}[1]{%
  \@ifundefined{alias@\string#1}
    {\ORIGselectlanguage{#1}}
    {\begingroup\edef\x{\endgroup
       \noexpand\ORIGselectlanguage{\@nameuse{alias@#1}}}\x}%
}
\newcommand{\definelanguagealias}[2]{%
  \@namedef{alias@#1}{#2}%
}
\makeatother
\definelanguagealias{en}{english}
\definelanguagealias{English}{english}
\usepackage{graphicx}
\usepackage{amsmath}
\usepackage{amsfonts}
\usepackage{amsthm}   
\usepackage{amssymb}
\usepackage{bm}
\usepackage{color}
\usepackage[percent]{overpic}
\usepackage{soul} 
\usepackage{amssymb}
\usepackage{wasysym}
\usepackage{dsfont}
\usepackage{float}

\usepackage{hyperref}
\hypersetup{
    bookmarks=false,         
    unicode=false,          
    pdftoolbar=false,        
    pdfmenubar=true,        
    pdffitwindow=false,     
    pdfstartview={FitH},    
    pdftitle={},    
    pdfauthor={Authors},     
    pdfsubject={},   
    pdfcreator={},   
    pdfproducer={}, 
    pdfkeywords={quantum many-body scars} {ergodicity}, 
    pdfnewwindow=true,      
    colorlinks=true,       
    linkcolor=black,          
    citecolor=blue,        
    filecolor=magenta,      
    urlcolor=blue           
}

\newcommand{\be}{\begin{equation}}
\newcommand{\ee}{\end{equation}}
\newcommand{\bea}{\begin{eqnarray}}
\newcommand{\eea}{\end{eqnarray}}

\usepackage{graphicx}
\usepackage[colorinlistoftodos]{todonotes}
\usepackage{verbatim}

\newcommand{\ket}[1]{\mbox{$| #1 \rangle$}}
\newcommand{\bra}[1]{\mbox{$\langle #1 |$}}

\usepackage{soul}
\usepackage[normalem]{ulem}

\newtheorem*{lemma*}{Lemma}

\begin{document}

\title{Emergent SU(2) dynamics and perfect quantum many-body scars}

\author{Soonwon Choi}
\thanks{S.\,C. and C.\,J.\,T contributed equally to this work.}
\affiliation{Department of Physics, University of California Berkeley, Berkeley, California 94720, USA}
\author{Christopher J. Turner}
\thanks{S.\,C. and C.\,J.\,T contributed equally to this work.}
\affiliation{School of Physics and Astronomy, University of Leeds, Leeds LS2 9JT, United Kingdom}
\author{Hannes Pichler}
\affiliation{Department of Physics, Harvard University, Cambridge, Massachusetts 02138, USA}
\affiliation{ITAMP, Harvard-Smithsonian Center for Astrophysics, Cambridge, MA 02138, USA}
\author{Wen~Wei Ho}
\affiliation{Department of Physics, Harvard University, Cambridge, Massachusetts 02138, USA}
\author{Alexios~A.~Michailidis}
\affiliation{IST Austria, Am Campus 1, 3400 Klosterneuburg, Austria}
\author{Zlatko Papi\'c}
\affiliation{School of Physics and Astronomy, University of Leeds, Leeds LS2 9JT, United Kingdom}
\author{Maksym Serbyn}
\affiliation{IST Austria, Am Campus 1, 3400 Klosterneuburg, Austria}
\author{Mikhail D. Lukin}
\affiliation{Department of Physics, Harvard University, Cambridge, Massachusetts 02138, USA}
\author{Dmitry A. Abanin}
\affiliation{Department of Theoretical Physics, University of Geneva, 1211 Geneva, Switzerland}

\begin{abstract}
Motivated by recent experimental observations of coherent many-body revivals in a constrained Rydberg atom chain, we construct a weak quasi-local deformation of the Rydberg blockade Hamiltonian, which makes the revivals virtually perfect. 
Our analysis suggests the existence of an underlying non-integrable Hamiltonian which supports an emergent SU(2)-spin dynamics within a small
 subspace of the many-body Hilbert space. 
We show that such perfect dynamics necessitates the existence of atypical, nonergodic energy eigenstates --- quantum many-body scars.
Furthermore, using these insights, we construct a toy model that hosts exact quantum many-body scars, providing an intuitive explanation of their origin.
Our results offer specific routes to enhancing coherent many-body revivals, and provide a step towards establishing the stability of quantum many-body scars in the thermodynamic limit.
 \end{abstract}

\maketitle

Remarkable experimental advances have recently enabled studies of nonequilibrium dynamics of isolated, strongly interacting quantum systems~\cite{Bloch2012,ladd2010quantum,PolkovnikovRMP}.
In such systems, it is commonly believed that a generic state initialized 
far from equilibrium eventually thermalizes, whereupon any initial local information  becomes unrecoverable~\cite{DeutschETH,SrednickiETH,RigolNature}.
While this process of thermalization provides the basis of   statistical mechanics, it also poses challenges for building large-scale quantum devices. Hence, it is of fundamental interest to understand mechanisms to evade thermalization.  Two well-studied possibilities include many-body localization in strongly disordered systems, and fine-tuned integrable systems~\cite{Huse-rev,AbaninRev,sutherland2004beautiful}. 

Recently,  quench experiments with Rydberg atom arrays~\cite{Schauss2012,Labuhn2016,Bernien2017} have discovered non-thermalizing dynamics of a new kind~\cite{Bernien2017}. Initialized in a high-energy N\'eel state, the system exhibited unexpectedly long-lived, periodic revivals, failing to thermalize on experimentally accessible timescales; in contrast, other high-energy product states exhibited thermalizing dynamics consistent with conventional expectations. 

These surprising observations have stimulated strong theoretical interest~\cite{Turner2017,wenwei18TDVPscar,TurnerPRB,khemani18integrability,lin2018exact}. Ref.~\cite{Turner2017} showed that the oscillatory dynamics stems from a small number of exceptional, nonthermal many-body eigenstates which are embedded in a sea of   thermal eigenstates, that generically obey the eigenstate thermalization hypothesis (ETH)~\cite{DeutschETH,SrednickiETH,RigolNature}. These atypical, ergodicity-breaking eigenstates were named `quantum many-body scars' in analogy to quantum scars in single-particle quantum systems, which are similarly nonergodic wavefunctions that concentrate along the unstable, periodic trajectories of the counterpart classical system~\cite{Heller84}. Ref.~\cite{wenwei18TDVPscar} firmed up this analogy by showing that the long-lived many-body revivals were also closely related to an unstable periodic orbit in a  variational, ``semiclassical'' description of the quantum many-body dynamics. 

Despite much theoretical effort, several key questions regarding the nature of quantum many-body scars remain open. 
In particular, owing to the slow decay, the ultimate fate of the revivals at very long times in the thermodynamic limit is not fully understood.
Another outstanding challenge is to understand the physical mechanism protecting scars in the Rygberg chain and beyond.
Ref.~\cite{khemani18integrability} conjectured that the observed revivals can arise due to proximity to a putative integrable point, where the whole spectrum (not just scarred eigenstates) become nonthermal. In particular, Ref.~\cite{khemani18integrability} demonstrated the existence of a nontrivial deformation of the Rydberg blockade Hamiltonian that results in a substantial modification of the many-body level statistics, that could be interpreted as proximity to integrability.
Moreover, earlier works~\cite{Bernevig2017,BernevigEnt} have demonstrated the coexistence of ETH-violating states in a  generically ergodic spectrum, by explicitly constructing exact many-body eigenstates of non-integrable AKLT model that feature low entanglement at arbitrary energy densities

\begin{figure*}
\centering
\includegraphics[width=1\textwidth]{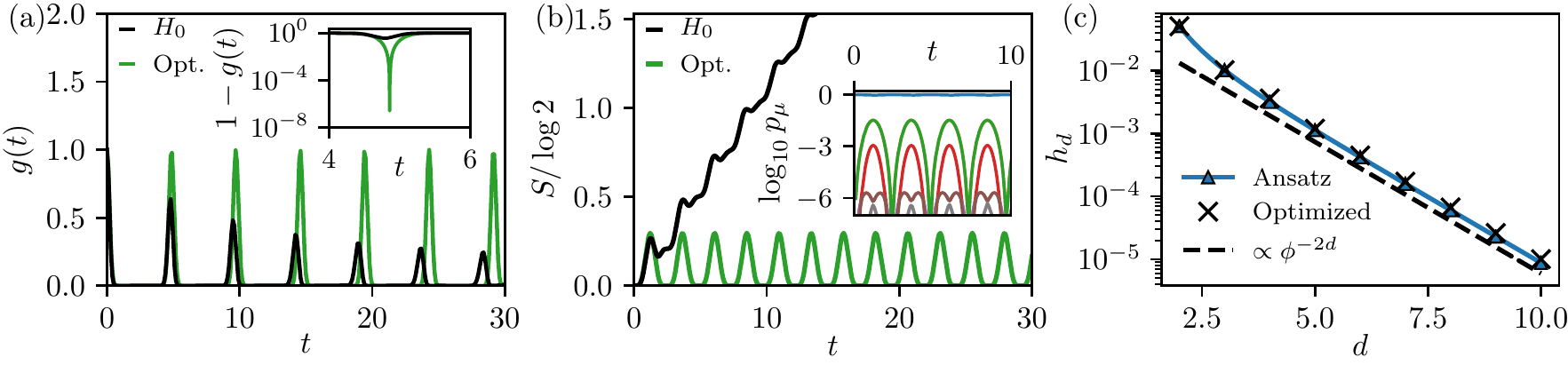}
\caption{ Non-thermalizing dynamics in  constrained spin Hamiltonians. 
(a) Many-body fidelity $g(t)$ as a function of time for the Hamiltonian $H_0$ without any perturbations and with optimal perturbations Eq.~(\ref{hk})-(\ref{eqn:ansatz}). The inset shows the infidelity, $1-g(t)$, on a logarithmic scale. (b) Half-chain bipartite entanglement entropy (EE) dynamics. At the optimal perturbation point, the EE as a function of time shows bounded, oscillatory dynamics. The inset shows the eigenvalues $p_\mu(t)$ of the half-chain reduced density matrix. Numerical simulations are performed with system size $N=32$ starting from the N\'eel state. (c) Optimized perturbation strengths $h_d$ decay exponentially. Solid line indicates the analytical ansatz function (\ref{eqn:ansatz}). } 
\label{fig01}
\end{figure*}

In the present work, we demonstrate that the periodic many-body revivals become extremely stable with a suitable weak, quasi-local deformation of the effective model describing the experiment~\cite{Bernien2017},  with the return probability of the N{\'e}el many-body state approaching unity within $10^{-6}$ in systems with more than 30 particles.
Remarkably, despite such manifestly nonergodic dynamics and the strongly nonthermal character of the associated scarred eigenstates, the bulk of the spectrum remains well-thermal, in contrast to the special point in Ref.~\cite{khemani18integrability}. 
Rather than being integrable, the revival dynamics can be understood as the coherent rotation of an emergent, large SU(2)-spin that lives within a special subspace of the many-body Hilbert space.

Our results strongly suggest the existence of a ``parent'' Hamiltonian with perfect oscillatory dynamics. We prove that, in generic settings, such perfect revivals impose strong constraints on the structure of energy eigenstates, necessitating the presence of some eigenstates violating ETH. This result directly relates observable nonequilibrium dynamics to properties of energy eigenstates, and parallels the mechanism behind quantum scarring in single-particle chaos theory~\cite{Heller84}. Finally, guided by the emergent SU(2)-spin structure, we construct a solvable toy model that explicitly hosts the phenomenology of quantum many-body scars, which provides an intuitive explanation of their origin in the constrained model.

\emph{Model and revivals.} -- The 1D chains of Rydberg atoms in the experiments~\cite{Bernien2017} are well-described by a kinetically constrained~\cite{Fredrickson1984,Palmer1984} spin-$1/2$ chain with the Hamiltonian
\begin{align}\label{eq:ham}
H_0 &= \sum_{i=1}^N  \mathcal{C} \sigma^x_i \mathcal{C},
\end{align}
where  $\sigma_i^\mu$ ($\mu\in \{x,y,z\}$) are   standard Pauli operators at site $i$, 
 and $\mathcal{C} = \prod_i[1- (1+\sigma^z_i)(1+\sigma^z_{i+1})/4]$ is a global projector
constraining the Hilbert space to spin configurations without two adjacent up-spins, $\ket{\uparrow \uparrow}$, corresponding to the regime of a strong nearest neighbor Rydberg blockade~\cite{jaksch2000fast} in the experiments~\cite{Bernien2017}. 
The dynamics is such that a spin may flip only when both of its neighbors  are in the $\ket{\downarrow}$ state, and the model is thus strongly interacting~~\cite{Sun2008,LesanovskyDynamics,Olmos2012}. 
For simplicity we assume periodic boundary conditions.

The model in Eq.~(\ref{eq:ham}) exhibits unexpected, long-lived periodic revivals when initialized in the N\'eel state $\ket{\mathbb{Z}_2} = \ket{\uparrow \downarrow \uparrow \downarrow \cdots}$. Despite its large energy density (formally corresponding to an infinite temperature), the quench dynamics from this initial state exhibits large recurrences  of the Loschmidt echo $g_0(t) \equiv |\bra{\mathbb{Z}_2}e^{-iH_0 t} \ket{\mathbb{Z}_2}|^2$ at multiples of a period $\tau$ with a slow overall decay (Fig.~1a) ~\cite{Bernien2017,Turner2017,wenwei18TDVPscar,TurnerPRB, khemani18integrability}. This is accompanied by a
generally linear growth of the  bipartite entanglement entropy (Fig.~1b), which is slower compared to other thermalizing initial states.  Such dynamics arise due to the existence of a band of special,  nonthermal `quantum many-body scarred' eigenstates that are approximately equally spaced in energy, and have large overlaps with $\ket{\mathbb{Z}_2}$. Furthermore, these special eigenstates can be approximately constructed using an analytical framework dubbed the forward scattering approximation (FSA)~\cite{Turner2017, TurnerPRB}. In essence, FSA relies on decomposing the Hamiltonian into a ``raising'' and ``lowering'' part, $H_0=H^+_0+H^-_0$, with $H^\pm_0 =  \sum_{i\in \text{even}}  \mathcal{C} \sigma_i^{\pm} \mathcal{C}+  \sum_{i\in \text{odd}}  \mathcal{C} \sigma_i^{\mp} \mathcal{C}$.
Then, one can recursively define $N+1$ vectors $|k\rangle_0 = \beta_k H^+_0 |k-1\rangle_0$ starting from $\ket{0}_0 = \ket{\mathbb{Z}_2}$, where $k\in\{0, 1, 2, \dots,N\}$ and $\beta_k$ is the normalization coefficient.
It has been shown that eigenstates belonging to the special band are predominantly supported by these FSA vectors spanning the subspace $\mathcal{K}$~\cite{Turner2017, TurnerPRB}.

\emph{Stabilizing revivals.}---In order to stabilize the revivals of $\ket{\mathbb{Z}_2}$, we considered various   perturbations that preserve the particle-hole and time-reversal symmetry of the system (thus, pinning the energy of $\ket{\mathbb{Z}_2}$). Generically, most peturbations weaken the revivals. However, we find that the following range-4 deformation 
\begin{align}\label{h2}
\delta H_2 = -\sum_i h_2 \mathcal{C} \sigma^x_i \mathcal{C} ( \sigma^z_{i+2} +  \sigma^z_{i-2})
\end{align}
with $h_2\approx 0.05$ (derived below), significantly improves the  fidelities of the revivals. 
We note that this form of perturbation has been previously considered in  Ref.~\cite{khemani18integrability}, which numerically found that at $h_2 \approx 0.024$, the entire spectrum becomes least thermal~\footnote{Our basis convention is different from Ref.~\cite{khemani18integrability}, which accounts for a trivial difference in sign of the optimal $h_2$.}. 
In contrast, our value of $h_2$ is approximately twice larger, and the spectrum remains thermal, aside from the  scarred eigenstates (see below).

Our key observation is that $\delta H_2$ partially cancels the errors arising in the FSA analysis.
More specifically, the precision of FSA, and therefore the stability and magnitude of revivals, relies on the dynamics of $\ket{\mathbb{Z}_2}$ generated by $H^\pm_0$ being (nearly) closed in the subspace $\mathcal{K}$. 
This condition would be exactly achieved if the vectors $|k\rangle$ were eigenstates of the operator $H^z_0\equiv[H_0^+,H_0^-]$, but is not  satisfied for $2\leq k \leq N-2$.
We find that this error can be reduced by adding $\delta H_2$ to the Hamiltonian and properly redefining the raising (lowing) operators, $H_2^\pm$, and the subspace $\mathcal{K}$ by replacing $\sigma_i^\pm \mapsto \sigma_i^\pm \left(1+h_2 (\sigma_{i+2}^z + \sigma_{i-2}^z)\right)$.
For example, one can analytically show that the component of $H_2^z \ket{2}$ perpendicular to $\ket{2}$ is minimized when $h_2 =1/2-1/\sqrt{5} \approx 0.053$~\cite{SOM}.
Surprisingly, this perturbation strongly improves many-body revivals, leading to fidelity $g(\tau) \approx 0.998$ at its first maximum for a system size $N = 32$. Furthermore, this deformation significantly slows down the growth of bipartite entanglement entropy~\cite{SOM}.

The dramatic increase in revival fidelities owing to $\delta H_2$ suggests that it might be possible to further enhance the oscillations, making them perfect. Extending our analytical considerations, it is natural to consider longer-range perturbations of the form
\begin{equation}\label{hk}
\delta H_R = -\sum_i \sum_{d =2}^R  h_d  \mathcal{C} \sigma^x_i  \mathcal{C} \left( \sigma^z_{i-d} + \sigma^z_{i+d} \right),
\end{equation}
which describe additional interactions between pairs of spins separated by a distance $d$, with strengths $\{ h_d \}$. We numerically optimize $\{h_d\}$ by maximizing the fidelity $g(t)$ under $H = H_0+\delta H_R$ at its first revival, whose results are summarized in Fig.~\ref{fig01}c for a system size $N = 20$ with $R=10$.
In~\cite{SOM}, we show that qualitatively similar results are obtained from other optimization methods, e.g. minimizing errors in FSA, etc.
We find that the optimized $h_d$ decay exponentially at large $d$, and can be intriguingly very well approximated by the analytical expression
\begin{align}
\label{eqn:ansatz}
h_d^{\textrm{ansatz}} =  h_0 \left( \phi^{(d-1)} - \phi^{-(d-1)}\right)^{-2},
\end{align}
where $\phi = \left(1+\sqrt{5}\right)/2$ is the golden ratio, and $h_0$ is a single parameter determining the overall strength. Henceforth, we will use  $h_d$ from  Eq.~(\ref{eqn:ansatz}) truncated at the maximum distance $N/2$, which allows us to perform a meaningful finite-size scaling analysis. Numerical optimization of the ansatz yields $h_0 \approx 0.051$.
Below, we will derive this value from certain algebraic relations of $H^\pm, H^z$ within the subspace $\mathcal{K}$.

Dynamics under the Hamiltonian $H = H_0 + \delta H_R$   makes the $|\mathbb{Z}_2\rangle$ revivals even more stable, with $1-g(\tau)\approx10^{-6}$ for $N=32$ at the first revival (Fig.~1a).  Simultaneously, we observe that the linear growth of the bipartitle entanglement entropy is significantly reduced, and is barely discernible (Fig.~\ref{fig01}b).
The scaling analysis in~\cite{SOM} suggests that the average rate of local thermalization, defined by the decay of $g(t)^{1/N}$, at late times vanishes in the thermodynamic limit.

\begin{figure}[t!]
\centering
\includegraphics[width=0.99\linewidth]{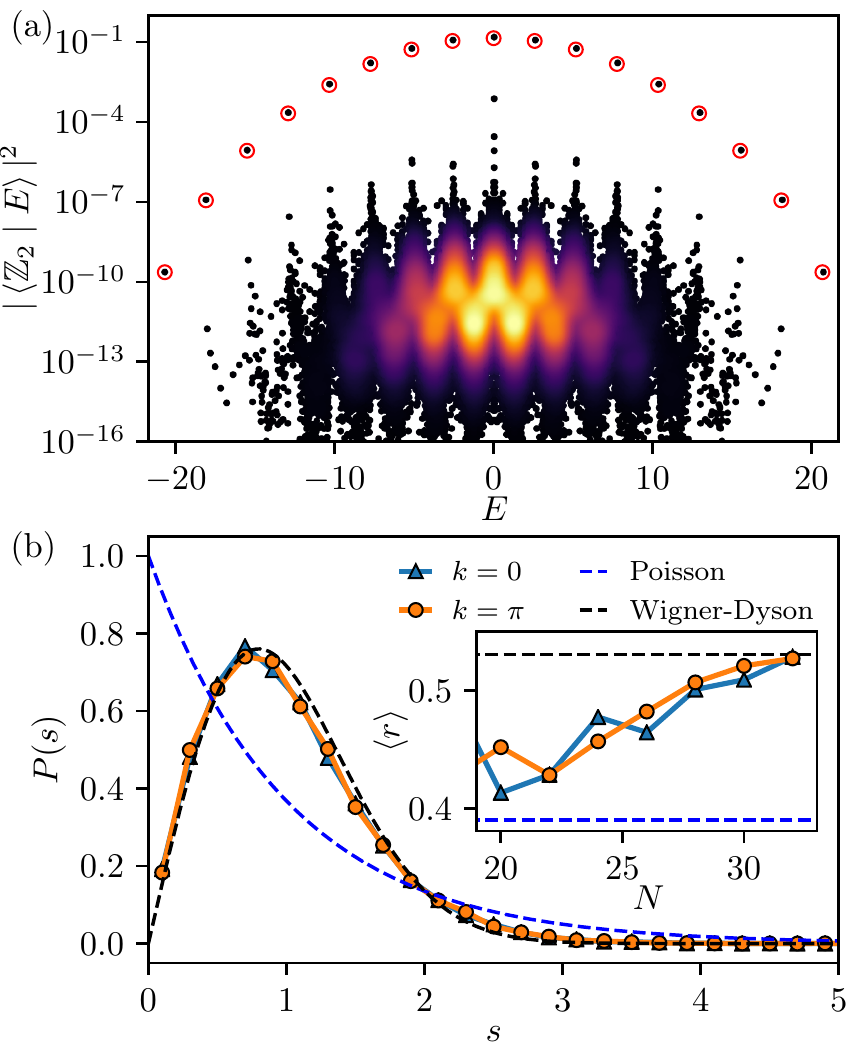}
\caption{ 
(a) Overlap of $|\mathbb{Z}_2\rangle$ with energy eigenbasis of $H$. 
The overlap is seen to be dominated by $N+1$ special eigenstates well separated from the bulk (highlighted by the red circles).
Data is shown for system size $N=32$ in the zero momentum and inversion symmetric sector.
(b) Eigenvalue level statistics closely follows that of Wigner-Dyson class of the Gaussian Orthogonal Ensemble. Inset: Level statistics indicator $\langle r_i \rangle$ as a function of system size $N$ flows to its value in the Wigner-Dyson ensemble, indicating that the bulk of the system remains ergodic.
Data shown is for system size $N=32$ in either the zero momentum and inversion even or the $\pi$ momentum and inversion odd sectors.
}
\label{fig02}
\end{figure}

{\it Dynamics constrains eigenstate properties.}---The  possible existence of a parent Hamiltonian leading to perfect oscillatory dynamics, strongly and quantifiably constrains the nonergodic nature of the quantum many-body scars. Specifically, we can appeal to the following general relation, whose proof is simple and given in~\cite{SOM}:

\emph{Lemma:
Consider a generic many-body Hamiltonian $H$ with extensive energy, $||H|| =O(N)$. If an initial state $\ket{\Psi_0}$ under time evolution perfectly comes back to itself after some time $\tau >0$, independent of the system size $N$, i.e. $|\bra{\Psi_0} e^{-iH\tau} \ket{\Psi_0}| =1$, then $\ket{\Psi_0}$ can be decomposed into $O(N)$ energy eigenstates, and at least one of them,  $\ket{\epsilon}$, has a large overlap $|\langle \epsilon | \Psi_0 \rangle|^2 \geq O(1/N)$. }

If the periodic revival occurs for a physical state $|\Psi_0\rangle$ with a finite energy density (that obey the cluster decomposition, so that the energy variance goes as $N$), such as $|\mathbb{Z}_2\rangle$ in our case, this Lemma dictates the presence of a high energy eigenstate with a large overlap $\sim 1/N$ with a low-entangled state. This constitutes a violation of the ergodic scenario, where a high energy eigenstate can be viewed as a random vector in the exponentially large Hilbert space.

In accordance with this result, the decomposition of the N\'eel state $|\mathbb{Z}_2\rangle$ in the energy eigenbasis of the perturbed Hamiltonian $H$ can seen to be dominated  by $N+1$ special  eigenstates  (Fig.~\ref{fig02}a), which are much better separated from the bulk than in the case of unperturbed Hamiltonian $H_0$. We also confirm that these eigenstates exhibit nonergodic properties, such as the logarithmic scaling of entanglement entropy, and can be constructed by a straightforward extension of FSA with significantly improved accuracy~\cite{TurnerPRB,SOM}.

Importantly, while the deformed model shows very stable revivals, the bulk of the spectrum remains thermal. To illustrate this, we compute the $r$-statistics associated to the level repulsion of the energy levels $E_i$,
$
\langle r_i \rangle = \left\langle {\textrm{min}(\delta_i, \delta_{i+1}) }/{ \textrm{max}(\delta_i , \delta_{i+1}) }\right\rangle,
$ 
where $\delta_i  = E_{i+1}- E_{i}$ is the level spacing and $\langle \cdot \rangle$ indicates averaging over a symmetry-resolved Hilbert space sector~\cite{footnote_symmetry_sectors}. 
Figure~\ref{fig02}b shows a clear flow in system size towards $\langle r_i \rangle \approx 0.53$, the Wigner-Dyson value associated with quantum chaotic Hamiltonians. In contrast, the Poisson level statistics  associated with the presence of integrable dynamics would correspond to $\langle r_i \rangle \approx 0.386$. We note that the flow of $\langle r_i \rangle$ towards its Wigner-Dyson value is faster than that of the unperturbed model $H_0$~\cite{Turner2017}, suggesting that the deformation enhances thermalization in the bulk.  
In addition, the probability distribution $P(s)$ of the unfolded level spacing $s$ is consistent with the Wigner-Dyson class of the Gaussian Orthogonal Ensemble.

\begin{figure}[tb]
  \centering
  \includegraphics{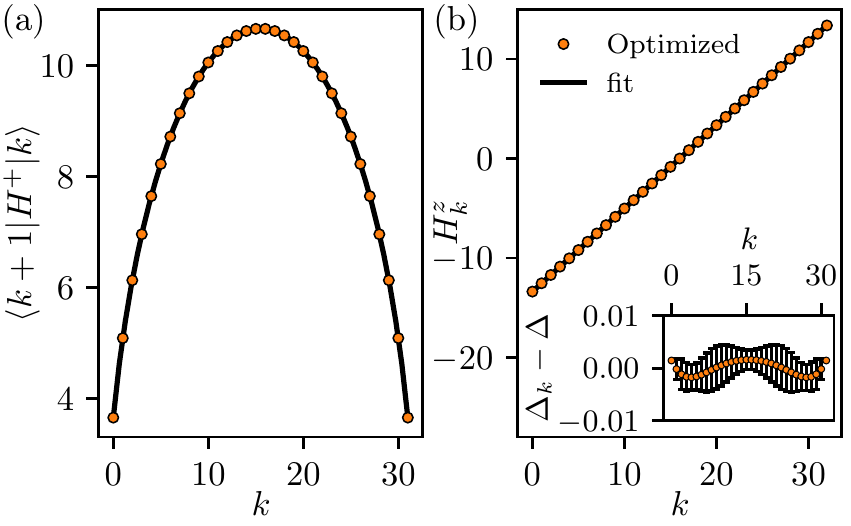}
  \caption{ \label{fig03}
 Emergent SU(2) structure in the subspace $\mathcal{K}$.
(a)  Matrix elements of the operator $H^+$ between consecutive vectors $\ket{k}$ are in excellent agreement with that of an appropriately rescaled raising operator $S^+$ in the $s = N/2$ representation of SU(2) shown as the solid curve.
(b) The FSA basis vectors $\ket{k}$ are approximate eigenstates of the operator $H^z$ with harmonically spaced eigenvalues. The inset shows the residual of the eigenvalue spacing  $\Delta_k \equiv \langle H^z \rangle_{k+1} -\langle H^z \rangle_{k}$ away from its mean value. 
The error bars are extracted from variances in the expectation values of $H^z$ for states $\ket{k}$.
}
\end{figure}

\emph{Algebraic structure in the subspace $\mathcal{K}$}.---The almost perfect fidelity revivals of the deformed Hamiltonian imply that operators $H^\pm$ and $H^z$ form a closed algebra within the subspace $\mathcal{K}$. Indeed we find numerically that
\begin{align}
\label{eqn:su2}
P_\mathcal{K} [H^z, H^\pm] P_\mathcal{K} \approx  \pm \Delta P_\mathcal{K} H^\pm P_\mathcal{K},
\end{align}
where $P_\mathcal{K} = \sum_k \ket{k}\bra{k}$   is the projector onto the subspace, and $\Delta$ is a constant. As $|0\rangle = |\mathbb{Z}_2\rangle$ is an eigenstate of $H^z$, $|k\rangle$ are also approximate eigenvectors of $H^z$ with harmonically spaced eigenvalues $H_k^z = \langle k | H^z | k \rangle$ so that $\Delta = H^z_{k+1} - H^z_k$. Thus, upon a suitable rescaling, the operator $H^z$ plays the role of $S^z$ in the SU(2) algebra, and $H^\pm$ play the role of spin-raising and lowering operators within $\cal K$. As the dimensionality of the subspace $\cal K$ is $N+1$, this implies that the operators form a spin $s = N/2$ representation of the SU(2) algebra, with $|\mathbb{Z}_2\rangle$ and $|{\mathbb{Z}'_2}\rangle = \ket{\downarrow \uparrow \downarrow \uparrow\dots}$ being the lowest and highest weight states respectively. To check this, we explicitly evaluated the matrix elements $\bra{k+1} H^+\ket{k}$. Figure~\ref{fig03}a confirms that up to an overall multiplicative factor, these matrix elements of $H^+$    reproduce the corresponding matrix elements of the spin raising operator $S^+$ in this representation. 

Thus, the virtually perfect oscillatory dynamics of $|\mathbb{Z}_2\rangle$ can be understood as a large spin $(s= N/2)$ pointing initially in an emergent ``$z$-direction'', undergoing a coherent Rabi oscillation under the Hamiltonian $H = H^+ + H^-$, which is akin to the $S^x$ operator, with period $\tau  = 2\pi/\sqrt{2\Delta}$. We stress that the emergence of this SU(2) structure within $\mathcal{K}$ is nontrivial, since the Hamiltonian $H$  by itself does not have any rotational symmetry.

The identification of this emergent algebra allows us to fix $h_0$ of our ansatz for $h_d$ analytically. In particular,  $H^z_k$ can be explicitly calculated for $k =0, 1$ in the thermodynamic limit. Imposing a harmonic spacing, i.e. $H^z_k = \Delta (k-N/2)$,  leads to a nontrivial constraint~\cite{SOM}
\begin{align}
\label{eqn:hd_constraint}
(1-h)(1-h- 16\sum_{n=1}^\infty h_{2n})= 16\sum_{n=1}^\infty h_{2n}^2,
\end{align}
where $h\equiv 2\sum_{n\geq 2} h_n (-1)^n$.
This fixes $h_0 \approx 0.0506656$ in our ansatz Eq.~(\ref{eqn:ansatz}), which agrees very well with the numerically optimized value. Furthermore, Eq.~\eqref{eqn:hd_constraint} determines the harmonic gap $\Delta = (1-h)^2\approx 0.835845$, and, correspondingly, the oscillation period $\tau\approx 4.85962$, which are also in excellent agreement with those from exact numerical simulations~\cite{SOM}. 

\emph{Toy model}.---The above investigations reveal that an emergent SU(2) structure within a special subspace underpins the   many-body revivals. Motivated by this, we construct a (solvable) toy model that exhibits similar phenomenology: in this model, there is a band of  nonthermal eigenstates supporting perfect oscillatory dynamics and exhibiting logarithmic entanglement, embedded in an otherwise thermal spectrum. 

Consider a system of $N$ spin-1/2 particles on a ring. The special subspace $\mathcal{V}$ of our model is defined as the common null space of $N$   projection operators $P_{i,i+1} = (1  - \vec{\sigma}_i \cdot \vec{\sigma}_{i+1}) /4$ onto neighboring pairs of singlets. Such subspace is spanned by the $N+1$ states of the largest spin representation $s=N/2$ of the SU(2) algebra. We enumerate the basis states of $\mathcal V$ by eigenstates of the $S^x =  \sum_i \sigma_i^x/2$ operator, $\ket{s=N/2, S^x =m_x}$ with $m_x \in \{-s, \dots s\}$.

Now, we take any Hamiltonian of the form 
\begin{align}\label{Eq:toy}
H_\text{toy} = \frac{\Omega}{2} \sum_i \sigma_i^x + \sum_i V_{i-1,i+2} P_{i,i+1},
\end{align}
 where $V_{ij}$ is a generic two-spin operator acting on spins $i$ and $j$, e.~g.~$V_{i,j} =  \sum_{\mu \nu}  J_{ij}^{\mu \nu} \sigma_i^\mu \sigma_j^\nu $  with arbitrary coefficients $J_{ij}^{\mu \nu}$. Note that $H_\text{toy}$ does not commute with $P_{i,i+1}$ nor $S^x$; thus, it does not have any obvious local symmetries. However, it can be easily verified that the   states $\ket{s=N/2, S^x =m_x} \in \cal V$  are eigenstates with harmonically spaced energies $E = \Omega m_x$. On the other hand, the states in the Hilbert space that do not belong to $\mathcal{V}$ are affected by the second term in Eq.~(\ref{Eq:toy}), and hybridize to form ergodic eigenstates~\cite{SOM}. Now, initializing our system, for example, in the lowest weight state $|N/2, S^z = -N/2\rangle$ leads to rotations of a large spin around the $x$-axis with frequency $\Omega$, whose motion remains in the subspace $\cal V$. We note that our construction is reminiscent   of Shiraishi and Mori's~\cite{ShiraishiMori} where a set of local projectors were used to embed  certain nonergodic energy eigenstates into the bulk of a many-body spectrum. 

Clearly, $H_\text{toy}$ exhibits all the features of perfect quantum many-body scarring, and is appealing as an intuitive understanding  of the origin of scars in the constrained spin models.
However, there remain many open questions: first,  the explicit relationship between the constrained spin model Eqs.~(\ref{eq:ham})-(\ref{hk}) and the toy model Eq.~(\ref{Eq:toy}) is not obvious. The nonisomorphic Hilbert spaces, as well as the nontrivial entanglement dynamics in the constrained model (Fig.~\ref{fig01}b), suggests that the mapping between these two models, if exists, cannot be strictly local. Second, it is highly desirable to find an analytic derivation of the deformation, Eq.~(\ref{hk}), that leads to the emergent SU(2) algebra  in the constrained spin model, and understand when such deformations exist for other local models. We note that this emergent algebra is reminiscent of the $\eta$-pairing symmetry that holds exactly in the Hubbard model~\cite{YangEta}, which allows to construct exact eigenstates at finite energy density with logarithmic~\cite{Vafek} and subthermal entanglement~\cite{Veness}. 
The exact expression for the analogue of $\eta$-pairing operator in our case, as well as the general relations between such operators, their algebra and scars, also remain an open question.

\emph{Summary and outlook}.--- To summarize, we have constructed a constrained spin model which exhibits nearly perfect quantum many-body scars. The remarkably long-lived oscillatory dynamics suggests that quantum scars remain stable in the thermodynamic limit. We showed that the dynamics can be understood in terms of a large, precessing SU(2) spin, and used this intuition to introduce a family of toy models with perfect scarring. In future work, it would be highly desirable to find an analytical mapping between the toy models and the constrained spin model. Moreover, the approach developed here may be applied to 
stabilize other types of quantum scars, in particular the ones originating from the $|\mathbb{Z}_3\rangle$ state in the model (\ref{eq:ham})~\cite{TurnerPRB}, as well as the ones found in higher-spin constrained models~\cite{wenwei18TDVPscar}. Another exciting challenge is to find models in which the MPS-based description of quantum scars trajectory becomes exact~\cite{wenwei18TDVPscar}. 
In a broader context, special non-thermalizing trajectories may have intriguing connections to  revivals/slow thermalization in strongly rotating gravitational systems~\cite{daSilva2015,DJafferis_chat}.
To understand the origin of this non-thermalizing dynamics, it would be valuable to establish whether quantum many-body scars can emerge from a dynamics that goes through states with high entanglement. 

\begin{acknowledgments}
{\it Acknowledgements.---}We thank E. Altman,  D. Jafferis, V. Khemani, S. Shenker, and especially Xiaoliang Qi for useful  discussions.
This work was supported through the National Science Foundation (NSF), the Center for Ultracold Atoms, the Air Force Office of Scientific Research via the MURI, and the Vannevar Bush Faculty Fellowship.
We are grateful to the KITP, which is supported by the National Science Foundation under Grant No. NSF PHY-1748958, and the program ``The Dynamics of Quantum Information'', where part of this work was completed.
S.C. acknowledges supports from the Miller Institute for Basic Research in Science.
 H.P. is supported by the NSF through a grant for the Institute for Theoretical Atomic, Molecular, and Optical Physics at Harvard University and the Smithsonian Astrophysical Observatory.
W.W.H. is supported by the Moore Foundation's EPiQS Initiative through Grant No.~GBMF4306.
D.A. acknowledges support by the Swiss NSF. 
C.J.T. and Z.P. acknowledge support by EPSRC grants EP/P009409/1, EP/R020612/1 and EP/M50807X/1. Statement of compliance with EPSRC policy framework on research data: This publication is theoretical work that does not require supporting research data.
\end{acknowledgments}
\bibliography{refs}

\begin{thebibliography}{35}
\bibitem [{foo()}]{footnote_symmetry_sectors}%
  \BibitemOpen
  \href@noop {} {}\bibinfo {note} {Since our Hamiltonian has spatial
  translation and inversion symmetries, we only diagonalize the symmetry sector
  with total momentum zero and even inversion parity. We exclude the degenerate
  energy eigenstates at zero energy since they originate from symmetry
  considerations.}\BibitemShut {Stop}%
\end{thebibliography}

\end{document}


\title{Supplementary Material for ``Emergent SU(2) dynamics and perfect quantum many-body scars"}
\date{\today}

\author{Soonwon Choi}
\thanks{S.\,C. and C.\,J.\,T contributed equally to this work.}
\affiliation{Department of Physics, University of California Berkeley, Berkeley, California 94720, USA}
\author{Christopher J. Turner}
\thanks{S.\,C. and C.\,J.\,T contributed equally to this work.}
\affiliation{School of Physics and Astronomy, University of Leeds, Leeds LS2 9JT, United Kingdom}
\author{Hannes Pichler}
\affiliation{Department of Physics, Harvard University, Cambridge, Massachusetts 02138, USA}
\affiliation{ITAMP, Harvard-Smithsonian Center for Astrophysics, Cambridge, MA 02138, USA}
\author{Wen~Wei Ho}
\affiliation{Department of Physics, Harvard University, Cambridge, Massachusetts 02138, USA}
\author{Alexios~A.~Michailidis}
\affiliation{IST Austria, Am Campus 1, 3400 Klosterneuburg, Austria}
\author{Zlatko Papi\'c}
\affiliation{School of Physics and Astronomy, University of Leeds, Leeds LS2 9JT, United Kingdom}
\author{Maksym Serbyn}
\affiliation{IST Austria, Am Campus 1, 3400 Klosterneuburg, Austria}
\author{Mikhail D. Lukin}
\affiliation{Department of Physics, Harvard University, Cambridge, Massachusetts 02138, USA}
\author{Dmitry A. Abanin}
\affiliation{Department of Theoretical Physics, University of Geneva, 1211 Geneva, Switzerland}


\maketitle

In this Supplementary Material, we provide detailed analytical and numerical analyses regarding the oscillatory dynamics and quantum many-body scars studied in the main text. In particular, we provide an analytical derivation of the optimal perturbation strength in the range-4 deformed model using the the framework of the forward scattering approximation (FSA). We then discuss and compare various numerical optimization schemes for determining the optimal couplings of the perturbed model. We also discuss finite-size scaling and fidelity dynamics for the quench from the N\'eel state, and furthermore provide detailed analysis of the eigenstate properties of the bulk spectrum, the special band of states, as well as the low-lying states (ground state and first excited state) of the deformed model. Finally, we prove the lemma in the main text, as well as show results of the numerical simulation of the toy model with exact quantum many-body scars.

\section{Optimal strength of the range-4 deformation}

In this Section we present the analytical derivation of the perturbation and its optimal magnitude. First we review the operator algebra and initial steps of the FSA in the constrained spin model Eq.~(1) from the main text (referred to as PXP model below). Afterwards, we derive the functional form of the deformation that minimizes the FSA error at the second step, and analytically estimate its magnitude. 

\subsection{Operator algebra and the FSA in the PXP model}\label{sec:fsa_range2}

We start with defining projected Pauli operators, distinguished by the tilde symbol,
\begin{equation}\label{Eq:}
\tilde{\sigma}^x_i=P_{i-1}\sigma^x_i P_{i+1}, 
\qquad 
 \tilde{\sigma}^z_i=P_{i-1}\sigma^z_i P_{i+1}, 
 \qquad \tilde{\sigma}_i^\pm=P_{i-1}\sigma_i^\pm P_{i+1}.
\end{equation}
This definition allows for a more compact representation of the PXP Hamiltonian, $H_0$, and its forward (backward) scattering parts. Namely,
\begin{equation}\label{Eq:}
H_0 = \sum_{i=1}^N \tilde{\sigma}^x_i =H_0^++H_0^-, 
\qquad 
H_0^{\pm}=\sum_{i=1}^{L} \left( \tilde{\sigma}_{2i-1}^{\mp} +\tilde{\sigma}_{2i}^{\pm} \right), 
\end{equation}
where we assume a spin chain of even size $N = 2L$ with periodic boundary conditions. The commutator between $H^\pm$ operators acts as an analog of the $S^z$ operator in the emergent SU(2) algebra (shown in the main text) and reads:
\begin{equation}\label{Eq:H^Z}
H_0^z = [H_0^+,H_0^-]
=
\sum_{i=1}^L \left( \tilde{\sigma}^z_{2i}-\tilde{\sigma}^z_{2i+1}\right).
\end{equation}
This operator is important, since below we show that the FSA condition that $H^-_0$ `inverts' the action of $H^+_0$ is equivalent to the condition that the FSA states are eigenstates of the operator $H^z_0$ with a certain eigenvalue. 

In order to demonstrate the role of the operator $H^z_0$, we start with considering the first three states of the FSA basis. 
The forward scattering basis is obtained via application of $H^+$ to the N\'eel state, $| k  \rangle  = \norm_k (H^+)^k|\mathbb{Z}_2\rangle$, where $\norm_k$ ensures the normalization, $\langle k|k \rangle=1$. The first three states read:
\begin{eqnarray}
|0\rangle &=& |\mathbb{Z}_2\rangle = |\uparrow\downarrow\uparrow\ldots\downarrow\uparrow\downarrow \rangle, \\
|1\rangle &=& \frac{1}{\sqrt{L}} \sum_{i=1}^{L} \tilde{\sigma}_{2i+1}^-|\mathbb{Z}_2\rangle, \\
 |2\rangle &=&\sqrt{\frac {1}{2L(L-1)}}  \sum_{i,j = 1; i\neq j}^L  \tilde{\sigma}_{2i+1}^- \tilde{\sigma}_{2j+1}^-|\mathbb{Z}_2\rangle.
\end{eqnarray}
The first FSA state, i.e., the N\'eel state itself, is an eigenstate of $H^z_0$. Using the explicit definition of this operator in Eq.~(\ref{Eq:H^Z}), we find that  $ H^z_0 |0\rangle = -L |0\rangle$, since operators $\tilde{\sigma}_i^z$ on even sites annihilate the $ |\mathbb{Z}_2\rangle$ state.

The condition $ H^z_0 |0\rangle = -L |0\rangle$ is equivalent to the FSA recurrence condition for $k=1$, 
\begin{equation}\label{Eq:FSA}
H^-_0 | k\rangle  = \frac{\norm_{k-1}}{\norm_k} |k-1\rangle.
\end{equation}
Indeed, the action of $H^-_0$ on the state $|1\rangle$ can be expressed as 
\begin{equation}\label{Eq:1check}
H_0^- | 1\rangle=\norm_1 H_0^- H_0^+ |0\rangle = \norm_1 (H_0^+ H_0^- -H^z_0)|0\rangle = - \norm_1 \langle H^z_0\rangle_0 |0\rangle,
\end{equation}
since $H_0^- |0\rangle$ annihilates $|0\rangle$ state and this state is an eigenvector of operator $H^z_0$ with the eigenvalue  $\langle H^z_0\rangle_0=-L$. Using the values of $\norm_0 = 1$, $\norm_1  = 1/\sqrt{L}$ one can confirm that Eq.~(\ref{Eq:1check}) implies that the condition~(\ref{Eq:FSA}) holds for $k=1$. 

Similarly, we can check that the state $|1\rangle$, which is a uniform superposition of spin flips at all possible positions respecting the constraint, is an eigenstate of $H^z_0$. The presence of one spin flip changes the expectation value of $H^z_0$ by $2$ compared to the N\'eel state, $ H^z_0 |1\rangle = (-L+2) |1\rangle$. This suffices to show that the FSA relation~(\ref{Eq:FSA}) holds for $k=2$:
\begin{equation}\label{Eq:2check}
H_0^- | 2\rangle=\norm_2 H_0^- [H_0^+]^2 |0\rangle
=
-\frac{\norm_2}{\norm_1} (  \langle H^z_0\rangle_0 + \langle H^z_0\rangle_1  ) | 1\rangle
  = \frac{\norm_{1}}{\norm_2} |1\rangle,
\end{equation}
since the first two eigenvalues of $H^z_0$ give us $ \langle H^z_0\rangle_0 + \langle H^z_0\rangle_1 = -2(L-1) = \norm_1^2/\norm_2^2$.

However, the FSA state $ |2\rangle$ is not an eigenstate of $H^z_0$, as this operator discriminates configurations with two next nearest neighbor spin flips,
\begin{equation}\label{Eq:Hz0-action2}
 H^z_0 |2\rangle  =  \sqrt{\frac {2}{L(L-1)}}  \left(
 (-L+3)\sum_{i = 1}^L  \tilde{\sigma}_{2i+1}^- \tilde{\sigma}_{2i+3}^-
+
(-L+4) \sum_{j>i = 1}^L  \tilde{\sigma}_{2i+1}^- \tilde{\sigma}_{2j+1}^-
 \right)|\mathbb{Z}_2\rangle .
\end{equation}
The different weight for configurations with two adjacent spin flips originates from the fact that the down spin at the site $2i+2$ also contributes to the expectation value of $H^z_0$ operator, whereas before it did not contribute due to the presence of projectors dressing the operator $\tilde{Z}_{2i+2}$. Since the FSA state $|2\rangle$ is not an eigenstate of $H^z_0$, the FSA stops being exact at the third step and the condition~(\ref{Eq:FSA}) no longer holds for $k=3$. 

While from the above we observe that the FSA state $ |2\rangle  $ is not an exact eigenstate of $H^z_0$, Eq.~(\ref{Eq:Hz0-action2}) suggests that the ``error" is small. Indeed, it would suffice to add terms $O(1/L)$ to make state $ |2\rangle  $ an eigenstate of $H^z_0$. Hence, this observation motivates us to seek a deformation of the Hamiltonian that improves the forward scattering beyond the first two steps.

\subsection{Deforming the Hamiltonian to correct for the first FSA error}

In order to find the deformation of the Hamiltonian, we use the intuition provided by Eq.~(\ref{Eq:Hz0-action2}). The action of $H^z_0$ on the FSA state $ |2\rangle  $ suggests that there exists an effective ``interaction" between spin flips that are 2 sites away from each other. Hence, in order to correct for such effective ``interaction" it is natural to consider the deformation of the Hamiltonian of the form $\delta H \propto \sum_{i} \tilde{\sigma}^x_i (P_{i-2}+P_{i+2})$ which corresponds to the increased rate of spin flips when there is a $\downarrow$-spin two sites away. Recalling that $1-2P_i = \sigma^z_i$, this deformation is equivalent to
\begin{eqnarray}\label{Eq:deform}
\delta H_2 = -h_2\sum_{i} \tilde{\sigma}^x_i (\sigma^z_{i-2}+\sigma^z_{i+2}), 
\end{eqnarray}
which was considered in the main text and also in Ref.~\cite{khemani18integrability}. Below we demonstrate that such a deformation allows to reduce the FSA error at the third step and estimate the optimal value of $h_2$ analytically. 

While we can use the intuition from FSA to write down the form of the deformation, calculating its effects requires reconsidering the operator algebra discussed above. In the presence of the deformation the splitting of the Hamiltonian into $H^\pm$ is modified as 
\begin{equation}\label{Eq:Hpm-deformed}
H^{\pm}=\sum_{i=1}^{L} \left( \tilde{\sigma}_{2i-1}^{\mp} W_{2i-1} +\tilde{\sigma}_{2i}^{\pm} W_{2i} \right),
\end{equation}
where we introduced the notation $W_i$ for the operator:
\begin{equation}\label{Eq:W-def}
W_i =\mathds{1}-h_2 (\sigma^z_{i-2}+ \sigma^z_{i+2}).
\end{equation}
Using this notation we write the operator $H^z$ as: 
\begin{equation}\label{Eq:Hz-deformed}
H^z = [H^+,H^-] =  \left(\sum_{i \in \text{even}}-\sum_{i \in \text{odd}}\right)\left( \tilde{\sigma}^z_{i} [W_{i}]^2 +2h_2  [2- h_2 (\sigma^z_{i-2}+\sigma^z_{i+4})][\tilde{\sigma}^-_{i}\tilde{\sigma}^+_{i+2}+\tilde{\sigma}^+_{i}\tilde{\sigma}^-_{i+2}]\right) 
\end{equation}
where we schematically indicate that sums over even and odd sites have opposite signs.

Finally, the FSA states $|0\rangle$ and $|1\rangle$ are not changed. However the eigenvalues of $H^z$ and normalization constants are now different: 
\begin{equation}\label{Eq:norms}
\norm_0= 1, \quad \langle H^z\rangle_0 = -L (1-2h_2)^2;
\quad 
\norm_1=\frac{1}{(1-2h_2)\sqrt{L}},
 \quad \langle H^z \rangle_1 = -L +2- 4h_2(1-h_2)(6-L). 
\end{equation}
In contrast, the state $ |2\rangle  $ that is generated by the action of $H^+$ operator from Eq.~(\ref{Eq:Hpm-deformed}) now takes a different form:
\begin{equation}\label{Eq:2-deformed}
|2\rangle =\norm_2  \sum_{i,j = 1; i\neq j}^L \tilde{\sigma}_{2i+1}^- W_{2i+1}  \tilde{\sigma}_{2j+1}^- W_{2j+1}|\mathbb{Z}_2\rangle
\\
=
2(1-2h_2) \norm_2  \sum_{i = 1}^L \left(
  \tilde{\sigma}_{2i+1}^- \tilde{\sigma}_{2i+3}^-
+
(1-2h_2) \sum_{j>i+1}  \tilde{\sigma}_{2i+1}^- \tilde{\sigma}_{2j+1}^-
 \right)|\mathbb{Z}_2\rangle,
\end{equation}
where the normalization $1/\norm_2 = 2(1-2h_2)\sqrt{L}\sqrt{1+(1-2h_2)^2(L-3)/2}$. Note that, as before, the conditions for $\langle H^z\rangle_0  = -\norm_0^2/\norm_1^2$ and $\langle H^z\rangle_0+\langle H^z\rangle_1  = -\norm_1^2/\norm_2^2$ hold. This implies that the deformation Eq.~(\ref{Eq:deform}) leaves exact the first two steps of the FSA for any value of $h_2$.

From Eq.~(\ref{Eq:2-deformed}) we see that the presence of $h_2\neq 0$ gives different weight to product with nearest neighbor spin flips compared to configurations with more distant spin flips. This was the source of error in the original PXP model. Hence, now we impose the condition that $|2\rangle$ is an eigenstate of $H^z$ and aim to find the optimal value of the perturbation strength, $h_2$. The action of  $H^z$ on $|2\rangle$ state reads:
\begin{equation}\label{Eq:2-deformed-Hz}
H^z|2\rangle =
2(1-2h_2) \norm_2  \sum_{i = 1}^L\left(
 f_2(h_2,L) \tilde{\sigma}_{2i+1}^- \tilde{\sigma}_{2i+3}^-
+
 f_4(h_2,L) \tilde{\sigma}_{2i+1}^- \tilde{\sigma}_{2i+5}^-
+
f_6(h_2,L) (1-2h_2) \sum_{j>i+2}  \tilde{\sigma}_{2i+1}^- \tilde{\sigma}_{2j+1}^-
 \right)|\mathbb{Z}_2\rangle.
\end{equation}
The evaluation of polynomials $f_i$ is cumbersome but straightforward, and results in 
\begin{eqnarray}\label{Eq:f}
 f_2(h,L) &=& -L+ 3 - 4 h(1-h)(7 -L),\\
 f_4(h,L) &=& -L+ 4 -2h[ 4 (7 - 14 h+ 8h^2 ) - (3 - 6h + 4h^2) L],\\
 f_6(h,L) &=& -L+ 4 -4 h (1 - h) (12 - L).
\end{eqnarray}
By comparing Eqs.~(\ref{Eq:2-deformed}) and (\ref{Eq:2-deformed-Hz}), we observe that  for $|2\rangle$ to be an eigenstate of $H^z$ it is necessary to have 
\begin{equation}\label{Eq:f2-f4}
 f_2(h^*_2,L) =   f_6(h^*_2,L) 
 \quad \to \quad
 1 - 20 h_2^* (1 - h_2^*) = 0.
\end{equation}
We note that all $L$-dependent terms cancel from this equation. Now, solving the resulting quadratic equation and choosing the smaller of the two solutions, we get the following expression for the optimal perturbation strength: 
\begin{equation}\label{Eq:h-opt}
h^*_2 = \frac{1}{2}- \frac{1}{\sqrt{5}} \approx 0.0527864,
\end{equation}
that agrees within the $5\%$ with the numerically determined perturbation strength listed in the main text. Moreover, we observe that the condition of the equidistant energy levels of operator $H^z$ implies:
\begin{equation}\label{Eq:}
 \langle H^z\rangle_1- \langle H^z\rangle_0 =  \langle H^z\rangle_2-\langle H^z\rangle_1.
\end{equation}
Using Eqs.~(\ref{Eq:norms}) and the fact that $\langle H^z\rangle_2 = f_2(h,L)$ we obtain the same condition as in Eq.~(\ref{Eq:h-opt}). This means that the optimal perturbation strength, $h_2^*$ results in \emph{both}, the state $|2\rangle$ being an approximate eigenstate of $H^z$ operator, \emph{and} in the first three eigenvalues of $H^z$ being equidistant from each other. 

We note, that the optimal value of $h^*_2$ determined above does not result in the state $|2\rangle $ being an exact eigenstate of $H^z$. The source of the remaining error are terms with the spin flips separated by \emph{four} sites. However, we observe that the error in the coefficient of the terms with spin flips 4 sites away is $L$-independent and is proportional to $h_2^2$. The direct calculation shows that it is small,
 \begin{equation}\label{Eq:4-sites-error}
 \epsilon  =  \Big[f_4(h_2,L)- (1-2h_2) f_6(h_2,L)\Big]_{h_2=h^*_2}
 =-32 [h^*_2]^2 (1 - h^*_2)
 \approx -0.0845.
\end{equation}
Including longer-range terms is expected to fix this error and also fix the magnitude of longer-range terms. We leave the detailed analytical investigation of the full hierarchy of perturbations to future work.

\section{Parameter optimization}

In the main text we presented results for a set of couplings $h_d$ optimized to maximize the return probability of the N\'eel state at the first revival.
We saw that this probability could be brought remarkably close to 1, without affecting thermalization dynamics of the rest of the system.
In this section we   discuss some details of the optimization performed in the main text, and compare that with alternative optimization schemes.
Since optimizing the fidelity revivals simultaneously tunes several properties of the system (e.g., the microscopic structure of eigenstates and their energy separation), we would like to understand the effect of perturbations on each one of these properties. For this we shall consider some alternative cost functions (or figures of merit) which capture different properties of the system that we expect to be important for producing the high return probability. As we explain below, these cost functions are defined to vanish for the optimal model when there is an exact SU(2) invariant subspace.

We start by introducing these different cost functions.
The first cost function is our main object of interest -- the return probability at the first revival. In a short time interval around each revival, the fidelity is reliably unimodal, and as such, we can use a golden-section search to efficiently determine the location of the revival peak (i.e., period of the revival).  On top of this, there is a variational optimization for the couplings $\{h_d\}$.  We do not have much intuition about the cost function expressed in terms of $\{h_d\}$, hence we resort to a Nelder-Mead simplex search, as implemented in Python Scipy package. We use the same type of search when optimizing any of the other cost functions mentioned below.

The second measure we consider is referred to as the \emph{subspace variance}.
This is defined as the sum of Ritz vector variances
\begin{equation}
  \trvar_\mathcal{K}(H) = \sum_{j=0}^{N} \var_{\tilde\psi_j} (H),
\end{equation}
where $\tilde\psi_j$ are the Ritz vectors, i.e., eigenstates of the projected Hamiltonian in the forward-scattering (FSA) subspace~\cite{Turner2017}.
This measures how well $\tilde\psi_j$ approximate the system's eigenstates.
The subspace variance can be rewritten as
\begin{align}
  \trvar_\mathcal{K}(H)
  &= \sum_j \bra{\tilde\psi_j} H^2 \ket{\tilde\psi_j} - \bra{\tilde\psi_j} H \ket{\tilde\psi_j} \bra{\tilde\psi_j} H \ket{\tilde\psi_j} =
 \sum_j \bra{\tilde\psi_j} \mathcal{K}(H^2) - (\mathcal{K}H)^2\ket{\tilde\psi_j}  \\ \nonumber 
  &= \tr\{\mathcal{K}(H^2) - \mathcal{K}(H)^2\}, \label{eq:subvar}
\end{align}
where $\mathcal{K}$ is the projector superoperator into the forward-scattering subspace.
In this way we can view it as a measure of how far $\mathcal{K}$ is from being a $H$-invariant subspace.
Note that if $\mathcal{K}$ were a one-dimensional subspace then the subspace variance reduces to the ordinary variance in that state.

We will use this idea to measure $H$-invariance when $\mathcal{K}$ is the forward-scattering subspace.
If the subspace variance were zero then the forward-scattering subspace would be closed, and the existence of such a low dimensional closed subspace is necessary for perfect revivals.
In the Lanczos basis, $\mathcal{K}(H)$ is a tridiagonal matrix with off-diagonal elements $\beta_1,\beta_2,\ldots,\beta_N$, hence $\tr\{(\mathcal{K} H)^2\} = 2 \sum_{j=1}^{N} \beta_j^2$.
The other term, $\tr\mathcal{K}(H^2) = \tr\{\mathcal{K}H^{+}H^{-} + \mathcal{K}H^{-}H^{+}\}$, since $H^{+}H^{+}$ is traceless.
Now, 
\begin{align}
  \tr\{\mathcal{K}H^{-}H^{+}\}
  &= \sum_{j=0}^{N} \bra{v_j} H^{-}H^{+} \ket{v_j}
  = \sum_{j=1}^{N} \beta_j^2,
\end{align}
and so we have that
\begin{equation}
  \trvar_\mathcal{K}(H) = \tr\{\mathcal{K}[H^{+},H^{-}]\} = \tr\{ \mathcal{K}H^z \}\text{.}
\end{equation}
The subspace variance is then the trace of the operator $H^z$ from the main text over the FSA subspace.

The third measure is a FSA error cost function which follows Section~\ref{sec:fsa_range2} in considering the first non-trivial forward-scattering error,
\begin{equation}
  \|\delta v_3 \|^2
  = \|H^{-} \ket{2} - \beta_{2} \ket{1}\|^2 
  = \bra{2}H^{+} H^{-}\ket{2} - \beta_{2}^2\text{.} \label{eq:fsaerr}
\end{equation}
This FSA error must be zero if the forward-scattering subspace is to be $H$-invariant, but is itself not a sufficient condition for the FSA subspace to be closed.

The final cost function we use seeks to measure how ``anharmonic" are the energy spacings.
We do this by taking the Ritz values (i.e.~the FSA approximation to the associated special eigenstates) and finding the least-squares fit to having equally spaced energy levels, i.e., minimizing the root-mean-square of the residuals.
This is consistent with the idea that in the effective decoupled subspace, the Hamiltonian acts as a $S^{x}$ generator of a large-spin representation of SU(2).
Unlike the previous cost functions which look at how well decoupled the forward-scattering subspace is from the rest of the system, the present Ritz-value cost function is sensitive to the dynamics within the subspace.

\begin{figure*}[htb]
\centering
\includegraphics[width=0.49\textwidth]{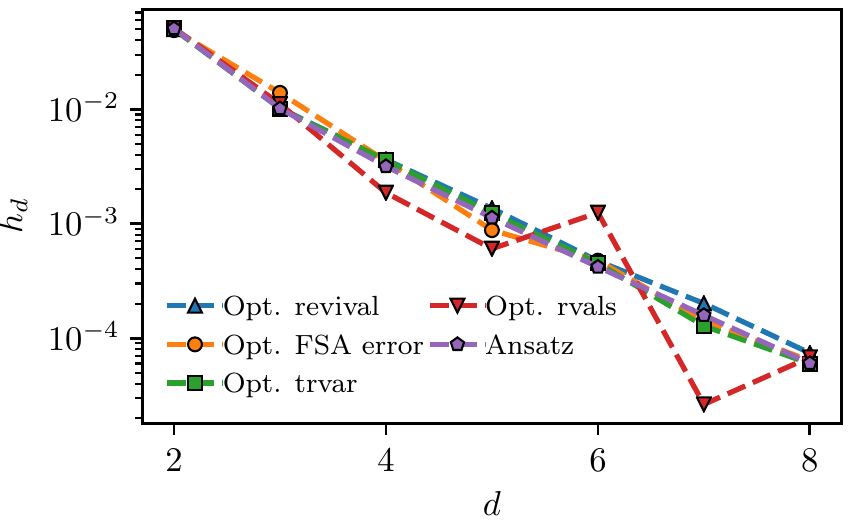}
\includegraphics[width=0.49\textwidth]{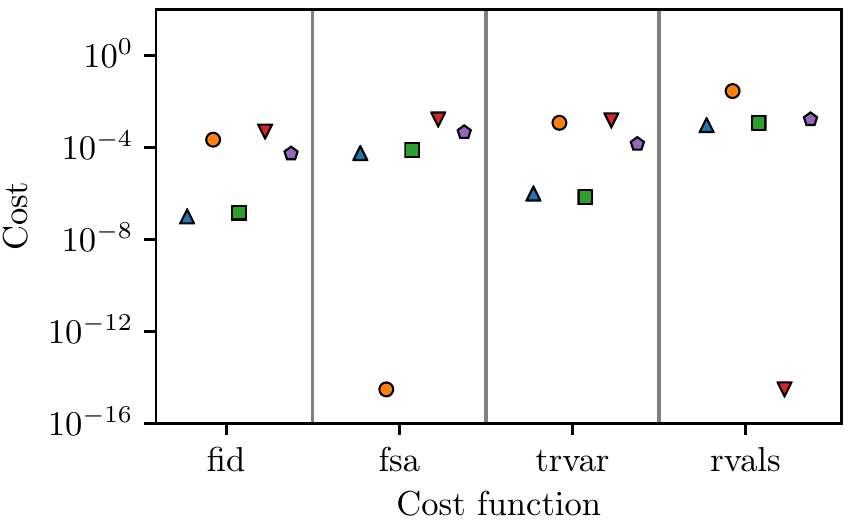}
\caption{
  (a) Coupling constants for the perturbation found by minimizing various cost functions.
  The labels:
  ``fid'' refers to the fidelity deficit at the first revival (also used in the main text);
  ``fsa'' to the first non-trivial forward-scattering error, Eq.~(\ref{eq:fsaerr});
  ``trvar'' to the subspace variance, Eq.~(\ref{eq:subvar}); and
  ``rvals'' to the anharmonicity of Ritz values. See text for details.
  (b) Evaluating the results of each of these optimization choices against these same figures of merit. (Different optimization choices are labelled in the same way as in the left panel.)}
\label{fig:S2}
\end{figure*}
In Figure~\ref{fig:S2}~(a), we show the coupling constants $h_d$ found numerically when optimizing each of these cost functions.
Different optimization schemes are found to result in approximately the same coupling constants $h_d$, and in particular they follow the same dependence on $d$ as the optimal ansatz presented in the main text. (Optimizing for harmonic spacing of the Ritz values leads to some non-monotonicity in the optimal coupling constants, which however still follow the main trend given by the ansatz.)

In Figure~\ref{fig:S2}~(b), we study the correlation between different optimization schemes, i.e., in each of the four panels we find the optimized coupling constants according to a particular cost function (denoted on the $x$-axis), and then evaluate the cost functions of other optimization schemes in order to assess their performance. One feature that particularly stands out is that optimizing for the first non-trivial error in the FSA is no longer sufficient to find the fidelity maximum once longer range terms are added beyond range 4. In contrast, optimizing for the subspace variance fares much better -- its optimal model also produces fidelity revivals with very similar accuracy. This highlights how the condition of having a low-dimensional approximately $H$-invariant subspace which contains an initial state is a stringent condition and strongly correlates with high-quality revivals.
Finally, while the harmonic spacing of the Ritz values is needed to produce revivals, it appears to trade off with increased line width. As a consequence, 
optimizing for harmonic energy spacing fares comparatively poorly in terms of return probability and the other figures of merit. In conclusion, being able to generate a low-dimensional, almost $H$-invariant subspace appears to be the most important factor responsible for the oscillatory dynamics in the fidelity.

\section{Constraint on perturbation strengths from SU(2) algebraic relations}
In this section, we provide a detailed explanation of how the effective SU(2) algebra discussed in the main text imposes a constraint on the perturbation strengths $\{h_d\}$, Eq.~(6), in the main text.
We first modify the definitions of $H^\pm$ and $H^z$ in a natural way: 
\begin{equation}\label{Eq:}
H^\pm =
 \sum_{i \in \text{even}} \tilde{\sigma}_i^\pm \left(1 - \sum_{d \geq 2} h_d (\sigma^z_{i+d} + \sigma^z_{i-d}) \right) + \sum_{i \in \text{odd}} \tilde{\sigma}_i^\mp \left(1 - \sum_{d \geq 2} h_d (\sigma^z_{i+d} + \sigma^z_{i-d}) \right),
\end{equation}
 and $H^z = H^+ H^- - H^- H^+$ as before.
Then, we will show that two states, $\ket{0} \equiv \ket{\mathbb{Z}_2}$ and $\ket{1} = \norm_1 \sum_{i \in \text{odd}} ( 1 - h) \sigma_i^- \ket{\mathbb{Z}_2}$ are exact eigenstates of $H^z$, where we introduced $h \equiv 2 \sum_{n\geq2} (-1)^n h_n$ for notational brevity.
 By explicit computation, it is easy show that 
 \begin{align}
H^z \ket{0} &= - H^- H^+ \ket{0} = -L (1 - h)^2 \ket{0}\\
H^z \ket{1} &= H^+ H^- \ket{1} - H^- H^+ \ket{1} =\norm_1 H^+ (H^- H^+ )\ket{0} - H^- H^+\ket{1}\\ \nonumber
&= \norm_1 H^+ \left( L (1-h)^2 \right) \ket{0} - H^- H^+ \ket{1}= \left( L (1-h)^2 \right) \ket{1} - H^- H^+ \ket{1},
\end{align}
where we again used the fact that we consider even system size written as $N=2L$. 
Thus, we see that $\ket{0}$ is an eigenstate of $H^z$, and $\ket{1}$ is also an eigenstate of $H^z$ as long as $H^- H^+ \ket{1} \propto \ket{1}$.
In fact, the latter is always the case:
\begin{align}
H^- H^+\ket{1} &= 4 \sum_{2\leq d\leq L,d \in \text{even} } (1 - h + 2h_{d})^2 \ket{1}= 4 \sum_{2\leq d\leq L,d \in \text{even} } \left[(1 - h)^2 + 4h_{d}(1-h) + 4 h_d^2 \right] \ket{1}\\ \nonumber
&= \left[(2L-2) (1-h)^2 + 16 \sum_{2\leq d \in \text{even} } h_{d} (1+h) +16 \sum_{2\leq d \in \text{even}} h_d^2 \right] \ket{1}.
\end{align}
This is because $\ket{1}$ is the unique state that is (i) invariant under the two-site translation and (ii) an eigenstate of the operator $\sum_{i\in \text{odd}} \tilde{\sigma}_i^z - \sum_{i \in \text{even}} \tilde{\sigma}_i^z$ with the eigenvalue $-2L+2 = -N+2$.
Note that we have extended the range of summation based on the assumption that $h_d$ decays sufficiently fast, e.g.~exponential as in our ansatz form.
Hence, we confirm that $\ket{0}$ and $\ket{1}$ are eigenvectors of $H^z$ with corresponding eigenvalues $-L (1-h)^2 $ and $ -\left( (L-2) (1-h)^2 + 16 \sum_{1\leq d  } h_{2d} (1-h) + 16\sum_{1\leq d } h_{2d}^2 \right)$ for an arbitrary choice of $\{h_d\}$.

Now, the effective SU(2) algebra among $H^\pm$ and $H^z$ (within the special subspace) requires that the eigenvalues of $H^z$ operator are harmonically spaced, as numerically verified in the main text.
Hence, these two values must satisfy the relation
\begin{align}
-L \Delta &= -L (1-h)^2 \\
-(L-1)  \Delta &= -L(1-h)^2 + 2(1-h)^2  - 16 \sum_{1\leq d  } h_{2d} (1-h) - 16\sum_{1\leq d } h_{2d}^2
\end{align}
for some $\Delta >0$.
This implies that
\begin{align}
(1-h)^2 - 16 \sum_{1\leq d  } h_{2d} (1-h) -16 \sum_{1\leq d } h_{2d}^2   = 0,
\end{align}
which provides a non-trivial constraint for $\{h_d\}$ discussed in the main text.
For our ansatz perburtation form, this constraint determines the overal strength $h_0 \approx 0.0506656$, and $\Delta \approx 0.835845$.

\section{Low lying special states in the deformed Hamiltonian}

In this Section, we consider the properties of low-lying special states. On   one hand, the low-lying special states play much smaller role in the revivals of the fidelity. On the other hand, the SU(2) symmetry realized on the manifold of the special states suggests that low lying eigenstates in this manifold also must have special properties. Below, we use DMRG for large system sizes to confirm that low lying states in the deformed Hamiltonian are also modified in a way consistent with the existence of SU(2) algebra. 

As was discussed previously,  the ground state $|\psi_0\rangle$ and the first excited state $|\psi_1\rangle$ of the PXP model are `special' eigenstates~\cite{TurnerPRB}. Hence, we consider how their properties change in the presence of the deformation for spin chains of size $N=60$. We use open boundary conditions and regularize the terms in the Hamiltonian near the boundary by removing the $\sigma^z$ operators that do not exist. For example, the first few terms in the Hamiltonian $H =\sum_{i=1}^N \mathfrak{h}_i$ near the boundary read:  
\begin{equation}\label{Eq:}
\mathfrak{h}_1 = \sigma^x_1 P_2  (1+\sum_{d= 2}^R h_d \sigma^z_{1+d}),
\quad 
\mathfrak{h}_2 = P_1\sigma^x_2 P_3  (1+\sum_{d= 2}^R h_d \sigma^z_{2+d}),
\quad
\mathfrak{h}_3 = P_2\sigma^x_3 P_4  (1+h_2\sigma^z_1 +\sum_{d= 2}^R h_d \sigma^z_{3+d}),
\quad \ldots.
\end{equation}
We target eigenstates using DMRG with projectors to the previous eigenstates with a large weight, e.g.\ we use $H_\text{eff}= H + w |\psi_0\rangle \langle \psi_0|$, where $|\psi_0\rangle$ is the ground state to find the first excited state. 
In order to represent the Hamiltonian, we use the matrix product operators with bond dimension $2 (4+R-1)$, where $R$ is the range of the interactions in the Hamiltonian. We have a fixed cutoff for singular values at $10^{-16}$, with bond dimension being adapted accordingly. 

\begin{figure}[t]
\begin{center}
\includegraphics[width=0.45\columnwidth]{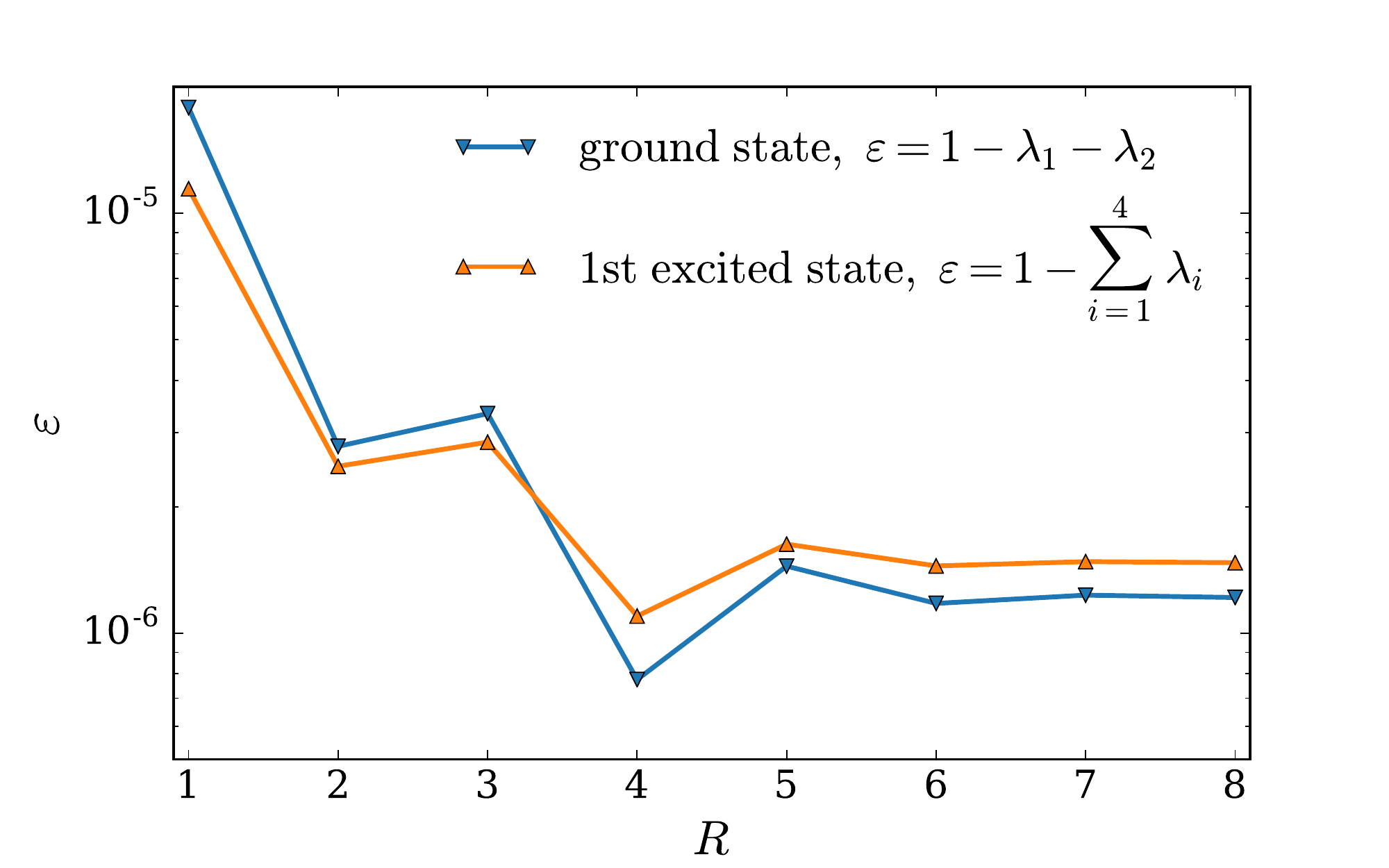}
\includegraphics[width=0.45\columnwidth]{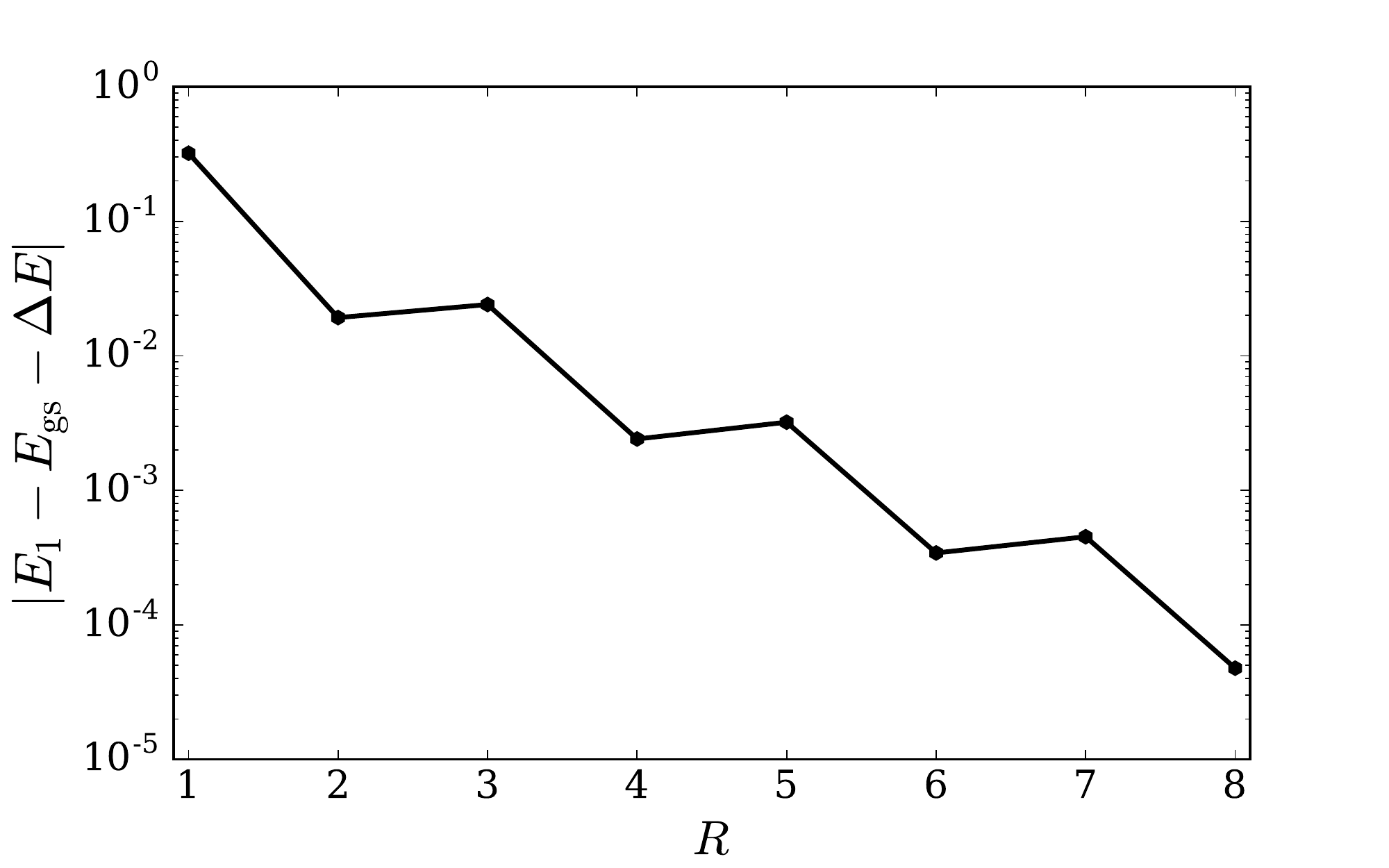}
\setlength{\unitlength}{\columnwidth}
\begin{picture}(0,0)
\put(-0.92, 0.25){(a)}
\put(-0.48, 0.25){(b)}
\end{picture}
\caption{ \label{Fig:gs}
Left panel shows that the ground state (the first excited state) with increasing perturbation range  are dominated by two (four) singular values.  $R=1$ corresponds to the original PXP Hamiltonian. Right panel illustrates the difference in energy between the first two special eigenstates, which rapidly approaches the theoretically predicted value with increasing $R$.}
\end{center}
\end{figure}

Figure~\ref{Fig:gs}(a) shows how the deformation of the Hamiltonian up to the given range $R$ enhances the leading singular values for the ground state and the first excited state. These singular values are calculated for the cut of the system in the middle to minimize boundary effects. In particular,  a perturbation of range $R=5$ leads to an order of magnitude decrease in the total weight contained by singular values $\lambda_i$ with $i\geq 3$ for the ground state. In other words, adding the deformation makes the ground state of the model to be more similar to the $\chi=2$ MPS state. Although the trend saturates for range $R>5$, we conjecture that this may be related to the finite size effects in optimization or to corrections on top of our perturbation ansatz.  Likewise, the second special eigenstate $|\psi_1\rangle$ seems to have dominant weight in the space spanned by the vectors associated to the first four singular values. These results suggest that the deformed model with perfect scars may have an exact bond-dimension 2 ground state.

Next, Figure~\ref{Fig:gs}(b) compares the energy difference between two states $|\psi_1\rangle$ and $|\psi_0\rangle$  to the prediction from SU(2) algebra, $\Delta E  = 2\pi/\tau \approx 1.29294$. We observe that the gap between numerically determined eigenstates approaches this value with finite size corrections decaying exponentially in $R$. Moreover, in contrast to Fig.~\ref{Fig:gs}(a) we see no sign of saturation up to the largest perturbation range $R=8$ considered here. 

\section{Scaling analysis of the oscillatory dynamics}

\begin{figure*}[tb]
  \centering
  \includegraphics[width=0.99\textwidth]{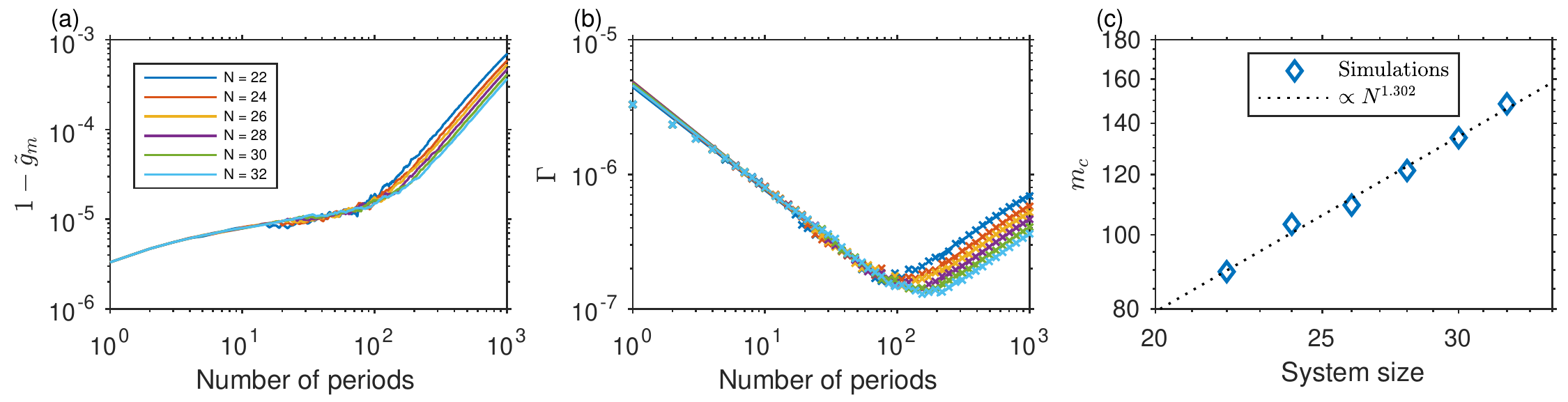}
  \caption{ \label{fig:scaling} 
  (a) Revival fidelities at late time. Lines with different color correspond to numerical simulations results for different system sizes. 
  (b) The average local thermalization rate $\Gamma$ decreases over time (in units of the revival period). The rates for different system sizes collapse on top of each other until they start growing after a critical time $m_c$. The value of $m_c$ can be extracted from the intersection between two independently fitted curves (solid lines) for $5\leq m \leq 60$ and for $200\leq m\leq 1000$.
  (c) The critical $m_c$ generally increases with system size $N$, suggesting that in the thermodynamic limit $m_c$ flows to infinity.  }
\end{figure*}

In this section, we provide numerical evidence that the local thermalization rate of the initial N\'eel state vanishes even in the thermodynamic limit via a finite-size scaling analysis.
We numerically evaluate the Loschmidt echo $g(t)$ under time evolution of the deformed Hamiltonian with the ansatz perturbation strengths in the main text for various system sizes $22\leq N\leq 32$.
We focus on the infidelity, $1-g$, at the $m$-th revivals at late time $1\ll m \leq 1000$.
In order to extract a meaningful finite-size scaling behavior, we first convert the infidelity to an intensive quantity that does not have explicit system size dependence. More specifically, we note that the overlap between two translationally invariant many-body wavefunctions generically decays exponentially with system size.
Hence, we normalize $\tilde{g} \equiv g^{1/N}$.
Heuristically, $\tilde{g}$ quantities the return probability per spin.
As shown in Fig.~\ref{fig:scaling}a, the normalized infidelity grows slowly (sub-linearly) for a certain number of periods, and then grows roughly quadratically.
More importantly, we find that $\tilde{g}$ does not depend on the system size and collapses on top of each other for relatively short times $m\lesssim100$, suggesting this behavior remains stable even in the thermodynamic limit.
At late times $m\gg 100$, the normalized infidelity \emph{decreases} with system sizes, implying that finite size effects are not negligible in this regime.
We envision that, in the thermodynamic limit, the short time regime extends to infinite times.
Note that in principle we cannot exclude the possibility that our ansatz for the perturbations is not sufficient, and one has to add more complicated multi-spin operators to the Hamiltonian to make the infidelity completely vanish.

Next we quantify the scaling behavior of $\tilde{g}$.
In particular, we are interested in how fast $\tilde{g}$ decays over long times.
Therefore, we introduce the average rate of decay (over one period) per spin :
 \begin{align}
\Gamma (m)
 \equiv \frac{1}{m} \left( 1 - \tilde{g}_m \right) 
=\frac{1}{m} \left( 1 - (1-(1-g_m))^{1/N} \right) 
\approx \frac{1-g_m}{Nm}, 
\end{align}
where $g_m$ is the Loschmidt echo at $m$-th revival and we used the fact that $1-g_m \ll 1$ in our regimes.
In our numerical calculations, we take the local maximum of $g(t)$ nearby $t = m \tau$ as a value of $g_m$.
As clearly seen in Fig.~\ref{fig:scaling}b, the average rates $\Gamma$ for different system sizes all collapse on top of each other and decrease up to a certain turning point $m_c$, after which they start increasing. The critical point $m_c$ increases with the system size $N$. 
In order to extract $m_c$ for each system size, we fit $\Gamma(m) = C m^\mu$ separately in two different regimes, short time ($5\leq m\leq 60$) and long time ($200\leq m\leq 1000$), and compute the intersection between the fitted curves.
As shown in Fig.~\ref{fig:scaling}c, the turning point $m_c$ increases with system sizes (slightly faster than linear).
These results suggest that, in the thermodynamic limit, $m_c$ diverges and $\Gamma$ approaches to zero indefinitely.

\section{Eigenstate ergodicity in the deformed model}

\begin{figure*}[htb]
  \centering
  \includegraphics[width=0.99\textwidth]{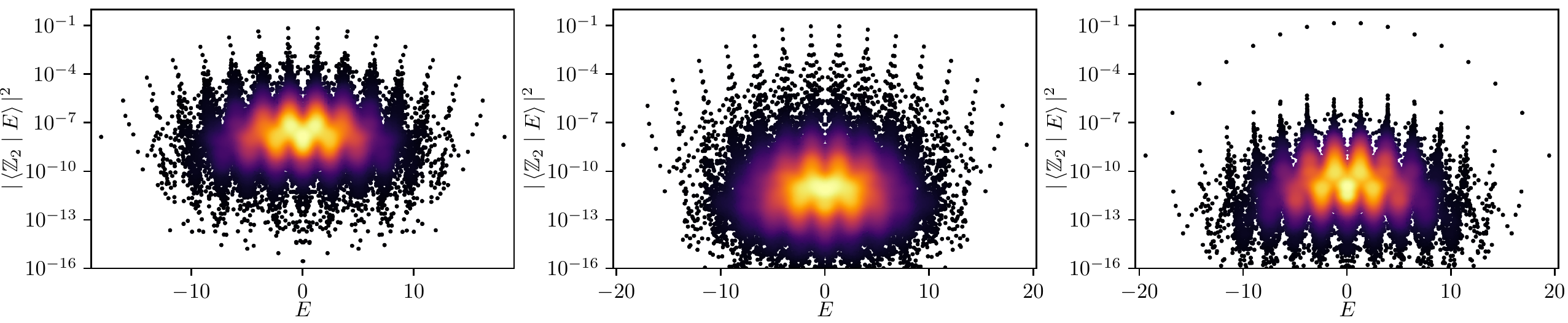}
  \includegraphics[width=0.99\textwidth]{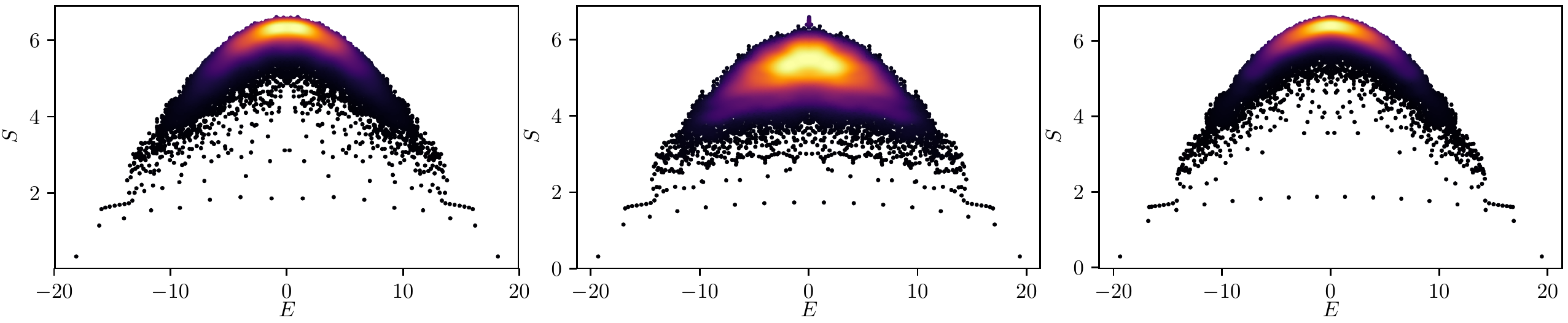}
  \caption{ \label{fig:eigs} 
    Top row: Overlap of all eigenstates with the $\mathbb{Z}_2$ product state plotted as a function of their energy. The three data sets correspond to the unperturbed PXP model (left), range-4 perturbation with strength $h_2 \approx 0.02$ (middle), and long-range ansatz from the main text (right). 
    Bottom row: Entanglement entropy of all eigenstates for a symmetric bipartition of the system, plotted as a function of their energy. The three data sets correspond to the same parameters as in the top row.    
    All data is for $N=30$ sites in the zero momentum and inversion symmetric sector.
  }
\end{figure*}

In this section we provide further data on the eigenstate properties of the deformed PXP model. It was previously established~\cite{Turner2017, TurnerPRB} that the scatterplot of the eigenstate overlap with the N\'eel state exhibits a characteristic fan structure, with a top band containing $N+1$ eigenstates, approximately equally spaced in energy, which maximize the overlap with the N\'eel state. This can also be seen in Fig.~\ref{fig:eigs}(top row, left) where we reproduce the same fan diagram for the unperturbed PXP chain with $N=30$ sites and periodic boundary conditions (zero momentum, inversion symmetric sector). These eigenstates underpin quantum revivals when the system is prepared in the N\'eel state. As noted in Ref.~\cite{TurnerPRB}, the special eigenstates responsible for the N\'eel revivals also form the band of lowest entanglement entropy states, see Fig.~\ref{fig:eigs}(bottom row, left). These special eigenstates, whose entropy is $S\lesssim 2$, are much more weakly entangled than the majority of eigenstates whose entropy is roughly $S \sim 6$ for this system size.

When the perturbation is turned on, the overlap diagram reorganizes in qualitatively different ways depending on the strength of the applied perturbation. In the middle panels of Fig.~\ref{fig:eigs}, we consider the range-4 perturbation with strength $h_2\approx 0.02$, i.e., the perturbation strength which was identified in Ref.~\cite{khemani18integrability} to correspond to a minimum in the level statistics parameter $r$. As we see in Fig.~\ref{fig:eigs}(top row, middle), the overlap towers are slightly broadened in this case, and the overlap of typical eigenstates with the N\'eel state is significantly reduced ($\sim 10^{-12}$ compared to $\sim 10^{-8}$ in the unperturbed PXP model). The main effect of the perturbation, however, is the broadening of entanglement entropy distribution in typical eigenstates and the reduction in average entanglement entropy in the system, see Fig.~\ref{fig:eigs}(bottom row, middle). This data is consistent with the slight enhancement of revivals and the deviation of the level statistics from Wigner-Dyson ensemble, as noted in Ref.~\cite{khemani18integrability}.

Finally, in right panels of Fig.~\ref{fig:eigs} we show the same results for the long-range perturbation introduced in the main text, with strengths fixed by the ansatz $h_d=h_0/(\phi^{d-1}-\phi^{-(d-1)})^2$. As we mentioned in the main text, the striking effect of this perturbation is that it (almost) fully  separates the band of $N+1$ special eigenstates from the rest of the states in the spectrum, in terms of the overlap with the N\'eel state, see Fig.~\ref{fig:eigs}(top panel, right). Moreover, while the overlap of typical eigenstates in the bulk of the spectrum is roughly similar to the case with $h_2\approx 0.02$ (middle panels), from the modulations in the density of states indicated by the color scale in Fig.~\ref{fig:eigs}, we see that clustering around the energies of special eigenstates is present throughout the spectrum (i.e., even amongst the eigenstates that have negligible overlaps $\sim 10^{-13}$ with the N\'eel state).

Remarkably, the enhanced scarring of eigenstates does not ``interfere" with the overall thermal properties of the rest of the bulk spectrum, diagnosed by the standard measures such as the average entanglement entropy. In fact, as we see in Fig.~\ref{fig:eigs}(bottom row, right), the average entanglement in eigenstates remains similar to the unperturbed PXP model, with the distribution of entropy concentrating even more sharply around the parabola which is typical of systems that obey the ETH. These results provide qualitative support of the claim in the main text that perfect scarring can coexist with the overall ergodic bulk spectrum, which was quantitatively established in the main text by evaluating the level statistics distribution. 

\begin{figure}[htb]
  \centering
  \includegraphics[width=0.4\linewidth]{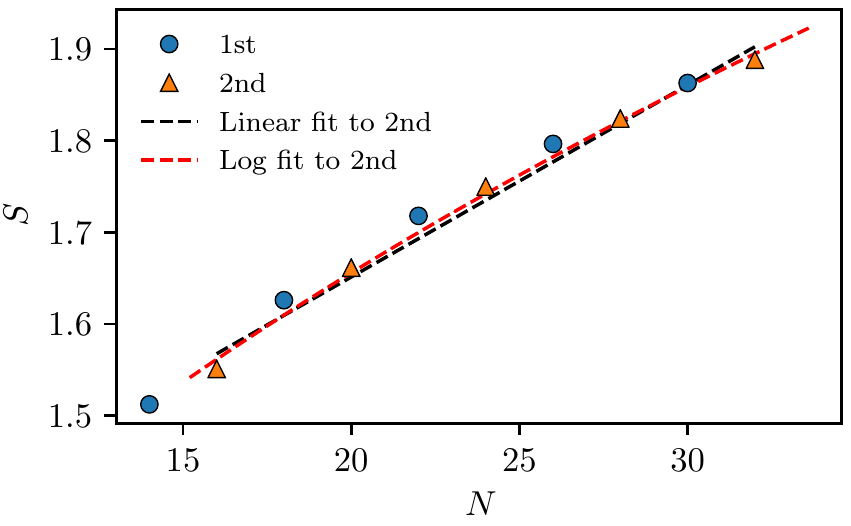}
  \caption{\label{fig:ent}
  Scaling of entanglement entropy as a function of system size in the model with optimal (ansatz) perturbations. Two sets of data points correspond to the first and second special eigenstate closest to the middle of the spectrum (energy $E=0$). For each set, we only show the data for system sizes where the corresponding state has zero momentum. Two dashed lines are linear and logarithmic fits to the data, respectively.
  }
\end{figure}
Finally, in Fig.~\ref{fig:ent} we study the scaling of the entanglement entropy with system size for the two special eigenstates closest to the middle of the spectrum. In Ref.~\cite{TurnerPRB} it was shown, within the FSA approximation, that special eigenstates have logarithmic scaling of entanglement entropy, $S\propto \ln N$. However, the entropy of \emph{exact} eigenstates behaved non-monotonically as a function of $N$, which was attributed to ``accidental hybridizations" with nearby eigenstates whose entropy scales with the volume of the subsystem~\cite{TurnerPRB}. In Fig.~\ref{fig:ent} we demonstrate that the optimal perturbation to the PXP model also suppresses the mixing with volume law states, at least up to the largest sizes accessible in exact diagonalization. Thus, the special eigenstates at the optimal point appear to  have logarithmic scaling of entanglement, $S\propto \ln N$, without volume law corrections.

\section{Proof of the Lemma}

In this section, we prove the Lemma in the main text.
We consider a generic Hamiltonian $H=\sum_{i=1}^N \mathfrak h_i$, where $N$ is the system size (volume) and each term $\mathfrak h_i$ has bounded norm $||\mathfrak h_i|| \leq   h$.
The bandwidth of $H$ scales at most linearly in $N$, i.e. $||H|| \leq N h$.
Here, we assume that there exists a certain initial state $\ket{\Psi_0}$ that has a perfect revival after unitary time evolution at some finite-time  $\tau$.
In terms of energy eigenstates $\{\ket{\mu}\}$,
\begin{align}
e^{-iH\tau} \ket{\Psi_0} &= e^{-iH\tau} \sum_\mu c_\mu \ket{\mu}
= \sum_\mu c_\mu e^{-i\epsilon_\mu \tau} \ket{\mu}
= e^{-i\alpha}  \sum_\mu c_\mu \ket{\mu},
\end{align}
where $c_\mu \equiv \langle \mu | \Psi_0\rangle$ and $\epsilon_\mu$ is the energy eigenvalue for $\ket{\mu}$.
The last equality holds owing to the perfect revival: $\bra{\Psi_0} e^{-iH\tau} \ket{\Psi_0} = e^{-i\alpha}$ for some $\alpha \in [0, 2\pi)$.
By comparing the last two expressions, we find that $c_\mu = 0$ unless $\epsilon_\mu = (\alpha + 2\pi m)/\tau $ for some integer $m$.
Without loss of generality, one may assume that $\ket{\Psi_0}$ has nonzero overlap with at most one energy eigenstate at each discrete energy $(\alpha + 2\pi m)/\tau$ because any linear combination of degenerate eigenstates is also an eigenstate.
Then, the bounded bandwidth of $H$ implies that there are at most $O(Nh \tau)$ energy eigenstates with non-vanishing overlap $c_\mu$.
In turn, we reach the conclusion that there must be at least one energy eigenstate with at least $O(1/\sqrt{N})$ overlap with the initial state $\ket{\Psi_0}$.

\section{Numerical results for the toy model}

In this section, we provide numerical simulation results regarding the toy model in the main text.
More specifically, we numerically diagonalize the Hamiltonian Eq.~(7) in the main text  for a system of $N = 14$ spin-1/2 particles under periodic boundary conditions.
We choose $\Omega = 1$ and $J^{\mu \nu}_{i-1,i+2}$ be independently drawn from a Gaussian distribution with zero mean and standard deviation $1/4$.
Note that, combined with the projectors $P_{i,i+1}$, this toy Hamiltonian describes a local spin chain with interactions up to range 4.
Generalizing this model to different spatial dimensions or geometry is straightforward by modifying $J_{ij}^{\mu \nu}$ appropriately.
In order to confirm that the model has exactly the same phenomenology of quantum many-body scarring, we compute three quantities: (i) the overlaps between a fully spin-polarized state $\ket{\psi_0}=\ket{\uparrow \uparrow \dots \uparrow}$ and energy eigenstates, (ii) Loschmidt echo $g(t) = |\langle \psi_0 | e^{-2\pi i H_\text{toy} t} | \psi_0 \rangle |^2$ for the quenched dynamics of $\ket{\psi_0}$, and the bipartite entanglement entropy of energy eigenstates.
\begin{figure}[htb]
  \centering
  \includegraphics[width=0.99\textwidth]{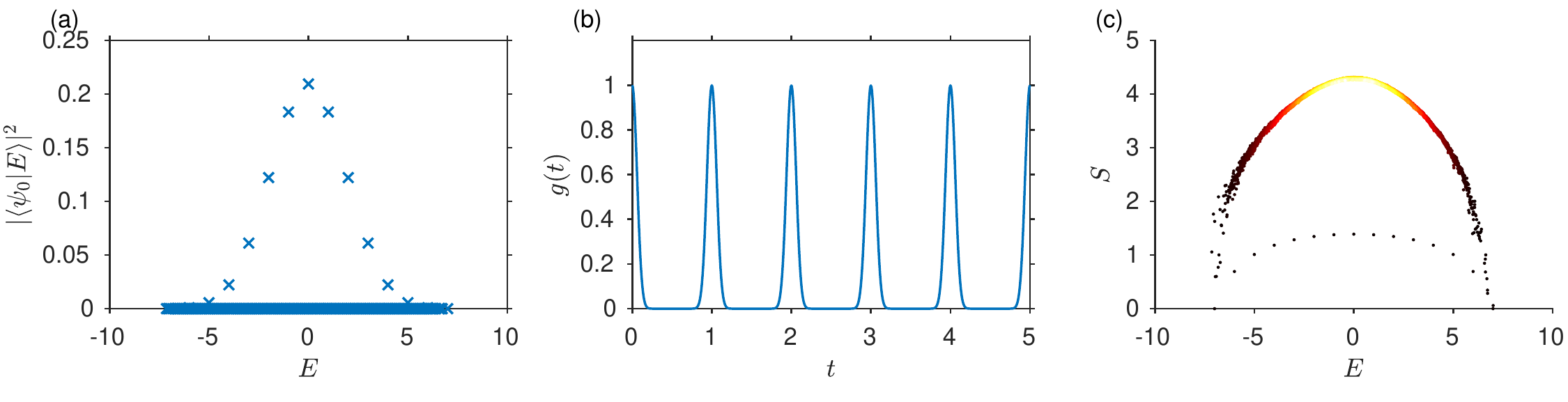}
  \caption{ \label{fig:toy_model} 
  (a) Overlaps between the spin-polarized state $\ket{\psi_0}=\ket{\uparrow \uparrow \dots \uparrow}$ and energy eigenstates for a system of $N = 14$ spin-1/2 particles.
  (b) Loschdmit echo $g(t)$ exhibits clear oscillatory behavior over period $\tau = 1$ without any decay, by construction.
  (c) Entanglement entropy of eigenstates as a function of energy $E$. Except for the special set of eigenstates, the rest of the eigenstates follows sharply the ETH predictions and have large entanglement entropies, indicating their ergodic nature. }
\end{figure}

As seen in Fig.~\ref{fig:toy_model}, all of these quantities clearly exhibit expected behavior, similar to the deformed PXP model of the constrained system. 
However, we stress once again that there are important differences between the toy model and the constrained model: in the former, perfect oscillatory dynamics of the fully spin-polarized state is accompanied with no entanglement dynamics whatsoever, while in the latter, the (almost) perfect oscillatory dynamics of the N\'eel state is accompanied with a small but nonzero oscillatory behavior of the entanglement entropy. Thus, a direct mapping between the two systems is possibly a nontrivial task. 
\bibliography{refs}